\documentclass[11pt,a4paper]{article}
\usepackage{amsmath,amsthm,amsfonts,amssymb,mathrsfs}
\usepackage{graphicx}
\usepackage{subfigure}
\usepackage[latin1]{inputenc}
\usepackage[swedish,english]{babel}
\DeclareMathOperator\sgn{sgn}
\renewcommand{\vec}[1]{{\boldsymbol#1}}
\allowdisplaybreaks

\begin{document}

\title{Determination of normalized electric eigenfields in microwave
    cavities with sharp edges}
\author{Johan Helsing\thanks{Centre for Mathematical Sciences, Lund
    University, Sweden}~~and Anders Karlsson\thanks{Electrical and
    Information Technology, Lund University, Sweden}}
\date{\today}
\maketitle
\begin{abstract}
  The magnetic field integral equation for axially symmetric cavities
  with perfectly conducting piecewise smooth surfaces is discretized
  according to a high-order convergent Fourier--Nyström scheme. The
  resulting solver is used to accurately determine eigenwavenumbers
  and normalized electric eigenfields in the entire computational
  domain.
\end{abstract}

\section{Introduction}

This work is on a numerical solver for the time harmonic Maxwell
equations in axially symmetric microwave cavities with piecewise
smooth and perfectly electric conducting (PEC) surfaces. We use the
interior magnetic field integral equation (MFIE) together with a
charge integral equation (ChIE) and high-order convergent
Fourier--Nyström discretization to find normalized electric
eigenfields to high accuracy.

The intended primary application of our solver is in computational
accelerator technology. Our experience is that our solver more than
doubles the range of frequencies for which electric and magnetic
eigenfields can be accurately evaluated, in comparison with finite
element method programs commonly used. This opens up for improved
evaluation of, so called, wakefields. Wakefields affect particle
trajectories during accelerator operation and wakefield prediction is
therefore of great importance in accelerator design, see~\cite[Chapter
11]{Wangler08}.

Our solver is based on the work~\cite{HelsKarl15}, which in turn draws
on progress in~\cite{BarnHass14,HelsKarl14,TaskOija06,Youn12}. A major
step forward from \cite{HelsKarl15} is the efficient treatment of
piecewise smooth surfaces and field singularities at sharp edges. For
this, we rely on a method called recursively compressed inverse
preconditioning (RCIP)~\cite{Hels08}.

The RCIP method can be seen as an automated tool to enhance the
performance of panel-based Nyström discretization schemes for Fredholm
second kind integral equations in the presence of boundary
singularities. For the determination of normalized eigenfields in
microwave cavities with sharp edges, and from a numerical point of
view, it is important to resolve boundary singularities and their
associated non-smooth fields to high precision. This is so since these
fields give non-negligible contributions to electric and magnetic
energies needed in the normalization. From a more practical point of
view in the accelerator design process, the identification and
evaluation of field singularities is necessary since strong fields may
cause field emission and quenching (thermal breakdown) in
superconducting cavities~\cite[Chapter 11 and 12]{Padam08}.

A common approach to the numerical resolution of fields at sharp edges
is to exploit a priori knowledge of asymptotic behavior and to include
a leading order singularity, or multiple non-integer powers, in
tailor-made basis functions. This approach generally reduces the
convergence order due to a dense spacing of presumptive
exponents~\cite{Bibbyetal08} and is difficult to automate and to apply
to problems that are not translationally invariant in one
direction~\cite{Idem00}. Still, it has been used in numerous papers
where the method of moments (MoM) is applied to scattering from PEC
structures with sharp edges~\cite{Bibbyetal08} and also to find stress
fields around cracks, notches, and grain boundary junctions in
computational mechanics~\cite{Link06}. The RCIP method, on the other
hand, does not require any known asymptotics and generally retains the
convergence order of the underlying discretization scheme.

The construction of our solver covers a wide range of topics and
computational techniques that are more or less well known and it would
carry too far to review them all. Some techniques that are considered
particularly important are discussed as they appear in the text. For
other issues we merely give references.

The outline of the paper is as follows: Section~\ref{sec:prob}
presents the MFIE and the ChIE for the problem at hand and an integral
representation for the electric field in a concise notation.
Section~\ref{sec:Fourier}, \ref{sec:normal}, \ref{sec:azimuthn}, and
\ref{sec:discrete} review the Fourier--Nyström discretization scheme
for smooth surfaces. The emphasis is on kernel evaluation and on the
conversion of a volume integral, used for normalization, into an
expression that is more suitable for numerics. Section~\ref{sec:RCIP}
is about sharp edges and how the RCIP method is incorporated into the
scheme. Here the ChIE plays an important role by simplifying the
accurate extraction of the surface charge density.
Section~\ref{sec:fysikalisk} relates the computed complex valued
electric fields to physical time-domain standing wave fields.
Section~\ref{sec:numerical} contains numerical examples with relevance
to accelerator technology and Section~\ref{sec:conclus} discusses
future work. In order to maintain a high narrative pace in the main
body of the paper, and also to provide an overview and to facilitate
coding, all explicit information on the integral operators used is
gathered in an appendix.

\section{Problem formulation} 
\label{sec:prob}

This section introduces the MFIE for the time harmonic Maxwell
equations in a notation that is particularly adapted to electric
fields inside axially symmetric cavities with PEC surfaces. Parts of
the material are well
known~\cite{Flemingetal05,GednMitt90,GlisWilt79,MautHarr69}. The
presentation parallels Section II of~\cite{HelsKarl15}, in which
magnetic fields are of primary interest.

\subsection{Geometry and unit vectors}
\label{sec:basic}

\begin{figure}
  \centering 
\noindent\makebox[\textwidth]{
\begin{minipage}{1.02\textwidth}
\hspace{1mm}
 \includegraphics[height=47mm]{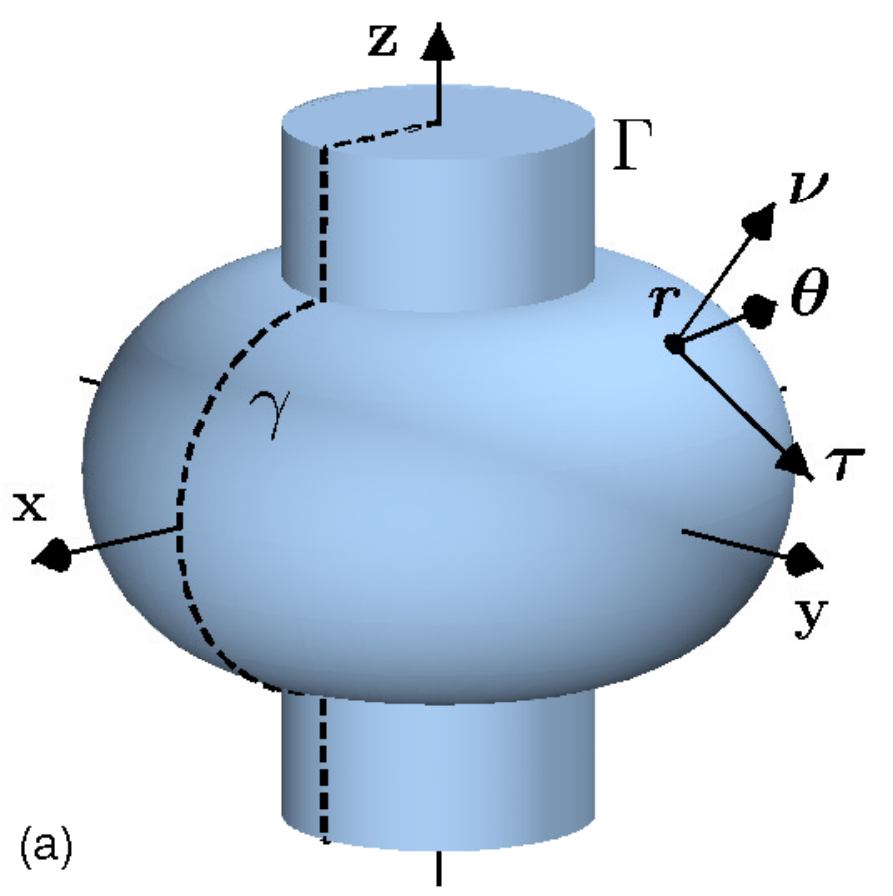}
 \includegraphics[height=47mm]{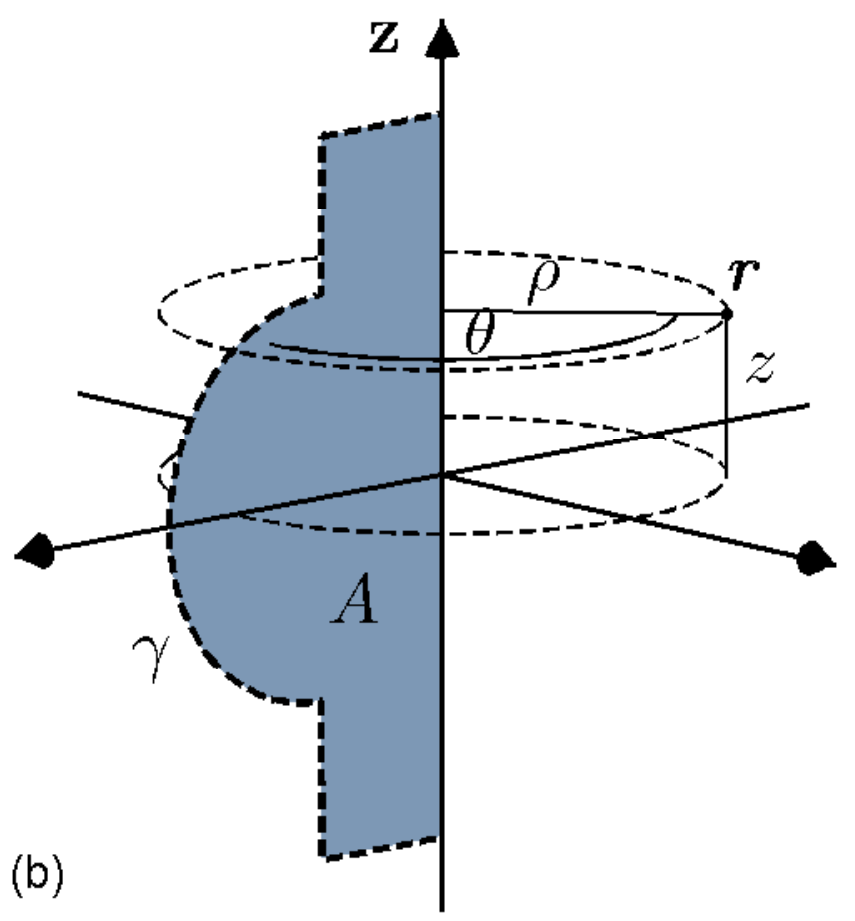}
 \includegraphics[height=47mm]{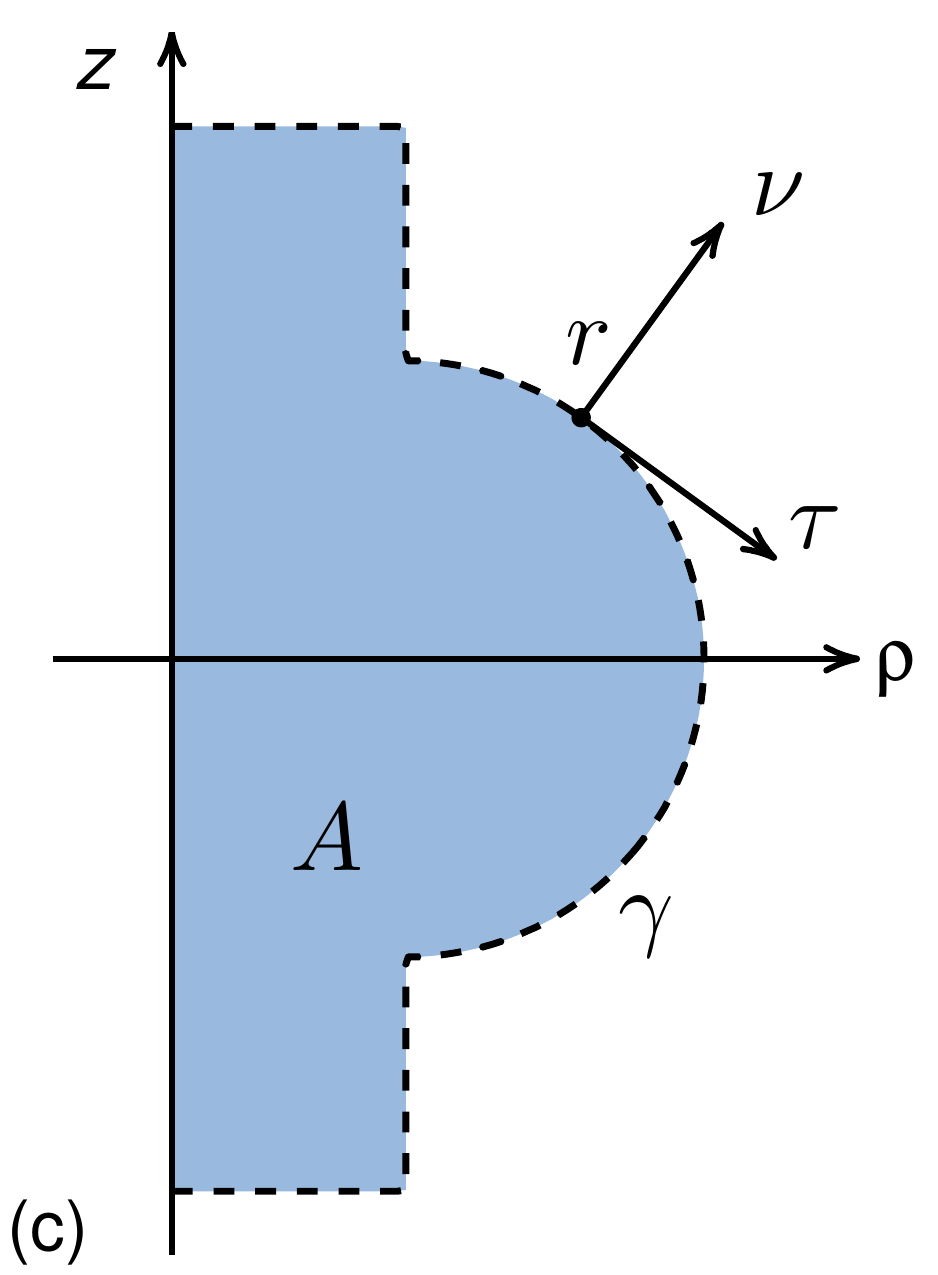}
\end{minipage}}
\caption{\sf An axisymmetric surface $\Gamma$ generated by a curve 
  $\gamma$. (a) A point ${\vec r}$ on $\Gamma$ has outward unit normal
  $\boldsymbol{\nu}$ and tangential vectors $\boldsymbol{\tau}$ and
  $\boldsymbol{\theta}$. (b) ${\vec r}$ has radial distance $\rho$,
  azimuthal angle $\theta$, and height $z$. The planar domain $A$ is
  bounded by $\gamma$ and the $z$-axis. (c) Coordinate axes and
  vectors in the half-plane $\theta=0$.}
\label{fig:geometry}
\end{figure}

Let $\Gamma$ be an axially symmetric surface enclosing a
three-dimensional domain $V$ (a body of revolution) and let
\begin{equation}
\vec r=(x,y,z)=(\rho\cos\theta,\rho\sin\theta,z)
\end{equation}
denote a point in $\mathbb{R}^3$. Here $\rho=\sqrt{x^2+y^2}$ is the
distance from the $z$-axis and $\theta$ is the azimuthal angle. The
outward unit normal $\vec\nu$ at a point $\vec r$ on $\Gamma$ is
\begin{equation}
\vec\nu=(\nu_{\rho }\cos\theta,\nu_{\rho }\sin\theta,\nu_z)\,.\\
\end{equation}
We also need the unit vectors
\begin{equation}
\begin{split}
\vec\rho&=(\cos\theta,\sin\theta,0)\,,\\
\vec\theta&=(-\sin\theta,\cos\theta,0)\,,\\
\vec\tau&=\vec\theta\times\vec\nu
 =(\nu_z\cos\theta, \nu_z\sin\theta,-\nu_{\rho })\,,\\
\vec z&=(0,0,1)\,,
\end{split}
\end{equation}
where $\vec\theta$ and $\vec\tau$ are tangential unit vectors.
See Figure~\ref{fig:geometry}(a) and~\ref{fig:geometry}(b).

The angle $\theta=0$ defines a half-plane in $\mathbb{R}^3$ whose
intersection with $\Gamma$ corresponds to a generating curve $\gamma$.
Let $r=(\rho,z)$ be a point in this half-plane and let $A$ be the
planar domain bounded by $\gamma$ and the $z$-axis. The outward unit
normal on $\gamma$ is $\nu=(\nu_\rho,\nu_z)$ and
$\tau=(\nu_z,-\nu_\rho)$ is a tangent. See
Figure~\ref{fig:geometry}(c). The unit vectors in the $\rho$- and
$z$-directions are $\hat{\rho}$ and $\hat{z}$.

\subsection{PDE formulation}

The electric field is scaled with the free space impedance $\eta_0$
such that $\vec E(\vec r)=\eta_0^{-1}\vec E'(\vec r)$, where $\vec
E'(\vec r)$ is the unscaled field. With vacuum in $V$ and with
$\Gamma$ perfectly conducting, the electric field $\vec E(\vec r)$
satisfies the system of partial differential equations
\begin{align}
\nabla^2\vec E(\vec r)+k^2\vec E(\vec r)&=\vec 0\,,\qquad\vec r\in V\,,
\label{eq:PDE1}\\
\nabla\cdot\vec E(\vec r)&=0\,,\qquad\vec r\in V\,,
\label{eq:PDE2}
\end{align}
with boundary condition
\begin{equation}
\lim_{V\ni \vec r\to \vec r^\circ}
\vec\nu^\circ\times\vec E(\vec r)=\vec 0\,, 
\qquad\vec r^\circ\in\Gamma\,.
\label{eq:BC}
\end{equation}
We will find nontrivial solutions to these equations in a fast and
accurate fashion via the MFIE. Note that, since $\Gamma$ is perfectly
conducting and from a mathematical as well as physical point of view,
nothing exterior to $\Gamma$ can affect $\vec E(\vec r)$ in $V$.
Hence the region exterior to $\Gamma$ is irrelevant to the problem.

The values $k^2$ for which the
system~(\ref{eq:PDE1}),~(\ref{eq:PDE2}), and~(\ref{eq:BC}) admits
nontrivial solutions are called eigenvalues. We refer to the
corresponding fields $\vec E(\vec r)$ as electric eigenfields and to
$k$ as eigenwavenumbers. The eigenvalues constitute a real, positive,
and countable set, accumulating only at infinity~\cite{Wen08}. The
eigenvalues have finite multiplicity. Electric eigenfields that
correspond to distinct eigenvalues are orthogonal with respect to the
inner product
\begin{equation}
\langle \vec F, \vec G\rangle=
\int_V \vec F^\ast(\vec r)\cdot\vec G(\vec r)\,{\rm d}V\,,
\label{eq:Hnormdef}
\end{equation}
where $\vec F(\vec r)$ and $\vec G(\vec r)$ are vector fields on $V$
and the asterisk indicates the complex conjugate. Subspaces of
electric eigenfields that correspond to degenerate eigenvalues can be
given orthogonal bases.

Electric eigenfields $\vec E(\vec r)$ are normalized so that
\begin{equation}
\|\vec E\|^2\equiv
\int_V\vec E^\ast(\vec r)\cdot\vec E(\vec r)\,{\rm d}V=1\,.
\label{eq:Enorm}
\end{equation}
The volume
integral in~(\ref{eq:Enorm}) is referred to as {\em the normalization
  integral}. In Section~\ref{sec:normal} it is converted into a
surface integral that is well suited for numerical evaluation in the
framework of the MFIE.

\subsection{Integral representation of the electric field}
\label{sec:elexp}

The surface current density $\vec J_{\rm s}$ and the surface charge
density $\varrho_{\rm s}$ are defined as
\begin{align}
\vec J_{\rm s}(\vec r^\circ)&=
\frac{{\rm i}}{k}\lim_{V\ni \vec r\to \vec r^\circ}
\vec\nu^\circ\times(\nabla\times \vec E(\vec r))\,, 
\qquad\vec r^\circ\in\Gamma\,,\\
\varrho_{\rm s}(\vec r^\circ)&=\lim_{V\ni \vec r\to \vec r^\circ}
-\vec\nu^\circ\cdot \vec E(\vec r)\,,\qquad\vec r^\circ\in\Gamma\,.
\label{eq:Varrho}
\end{align}
When $\vec J_{\rm s}$ and $\varrho_{\rm s}$ are known the electric
field is given by the integral representation
\begin{equation}
\left.\begin{aligned}
&\vec E(\vec r)\\
&\vec 0
\end{aligned}\right\}=
{\rm i} k\int_\Gamma
\vec J_{\rm s}(\vec r')\Phi_k(\vec r,\vec r')\,{\rm d}\Gamma'-
\int_\Gamma
\varrho_{\rm s}(\vec r')\nabla\Phi_k(\vec r,\vec r')\,{\rm d}\Gamma'
\left\{\begin{aligned}
&\vec r\in V\,,\\
&\vec r\in c\overline{V}\,,
\end{aligned}\right.
\label{eq:Erep}
\end{equation}
where $c\overline{V}$ is the exterior to $V\cup\Gamma$. Here, using
the time dependence $e^{-{\rm i}\omega t}$ with angular frequency
$\omega>0$, the kernel
\begin{equation}
\Phi_k(\vec r,\vec r')=\frac{e^{{\rm i}k\vert\vec r-\vec r'\vert}}
{4\pi\vert\vec r-\vec r'\vert}
\label{eq:funda}
\end{equation}
is the causal fundamental solution to the Helmholtz equation and
\begin{equation}
\nabla\Phi_k(\vec r,\vec r')=-\frac{(\vec r-\vec r')}
{4\pi\vert\vec r-\vec r'\vert^3}{P(\vert\vec r-\vec r'\vert)}\,,
\end{equation}
where
\begin{equation}
P(|\vec r-\vec r'|)=
(1-{\rm i}k\vert\vec r-\vec r'\vert)e^{{\rm i}k\vert\vec r-\vec r'\vert}\,.
\end{equation}
The lower equation in~(\ref{eq:Erep}) states that $\vec J_{\rm s}$ and
$\varrho_{\rm s}$ induce an electric null field outside $\Gamma$.

We decompose the electric field in its cylindrical coordinate
components
\begin{equation}
\vec E(\vec r)=\vec\rho E_{\rho}(\vec r)
+\vec\theta E_{\theta}(\vec r)+\vec zE_z(\vec r)\,
\end{equation}
and the surface current density in its tangential components
\begin{equation}
\vec J_{\rm s}(\vec r)=
\vec\tau J_{\tau}(\vec r)+\vec\theta J_{\theta}(\vec r)\,.
\end{equation}
From~(\ref{eq:Erep}) the components of the electric field, including
the induced external electric null field, can be expressed as
\begin{equation}
\begin{split}
E_{\rho}(\vec r)&={\rm i}kS_7J_\tau(\vec r)
                  +kS_8J_\theta(\vec r)+K_{11}\varrho_{\rm s}(\vec r)\,,\\
E_{\theta}(\vec r)&=kS_9J_\tau(\vec r)+{\rm i}kS_{10}J_\theta(\vec r)
                  +{\rm i}K_{12}\varrho_{\rm s}(\vec r)\,,\\
E_{z}(\vec r)&={\rm i}kS_{11}J_\tau(\vec r)+K_{13}\varrho_{\rm s}(\vec r)\,,
\end{split}
\label{eq:comp13}
\end{equation}
where $\vec r\in\mathbb{R}^3\setminus\Gamma$ and the double-layer type
operators $K_\alpha$ and single-layer type operators $S_\alpha$ with
various indices $\alpha$ are defined by their actions on a layer
density $g({\vec r})$, ${\vec r}\in\Gamma$, as
\begin{align}
K_\alpha g({\vec r})&=
\int_\Gamma K_\alpha(\vec r,\vec r')g({\vec r'})\,{\rm d}\Gamma'\,,
\label{eq:Kdefa}\\
K_\alpha(\vec r,\vec r')&=D_\alpha(\vec r,\vec r')P(|\vec r-\vec r'|)\,, 
\label{eq:Kdefb}
\end{align}
and
\begin{align}
S_\alpha g({\vec r})&=
\int_\Gamma S_\alpha(\vec r,\vec r')g({\vec r'})\,{\rm d}\Gamma'\,,
\label{eq:Sdefa}\\
S_\alpha(\vec r,\vec r')&=
Z_\alpha(\vec r,\vec r')e^{{\rm i}k\vert\vec r-\vec r'\vert}\,.
\label{eq:Sdefb}
\end{align}
The functions $D_\alpha(\vec r,\vec r')$ and $Z_\alpha(\vec r,\vec
r')$ can be viewed as static kernels, corresponding to wavenumber
$k=0$. Their explicit expressions are given in
Appendix~\ref{sec:explicitS}.

\subsection{Integral equations for $\vec J_{\rm s}$ and $\varrho_{\rm s}$}

The interior MFIE for $\vec J_{\rm s}$ reads
\begin{equation}
\vec J_{\rm s}(\vec r)-2\,\vec{\nu}\times\int_\Gamma\left(\vec J_{\rm s}(\vec
  r')\times\nabla \Phi_k(\vec r,\vec r')\right)\,{\rm d}\Gamma'=\vec 0\,,
\qquad \vec r\in \Gamma\,.
\label{eq:MFIE}
\end{equation}
The surface charge density $\varrho_{\rm s}$ is related to the surface
current density $\vec J_{\rm s}$ by the continuity equation
\begin{equation}
\varrho_{\rm s}(\vec r)=
-\frac{\rm i}{k}\nabla_{\rm s}\cdot\vec J_{\rm s}(\vec r)\,,
\label{eq:diffj}
\end{equation}
where $\nabla_{\rm s}\cdot(\;)$ is the surface divergence. As
in~\cite{HelsKarl15}, we avoid the differentiation inherent
in~(\ref{eq:diffj}) by evaluating $\varrho_{\rm s}$ from the interior
charge integral equation
\begin{multline}
\varrho_{\rm s}(\vec r)
-2\int_\Gamma\vec\nu \cdot\nabla\Phi_k(\vec r,\vec r')
\varrho_{\rm s}(\vec r')\,{\rm d}\Gamma'=\\
-2{\rm i}k\int_\Gamma\vec\nu\cdot\vec J_{\rm s}(\vec r')
\Phi_k(\vec r,\vec r')\,{\rm d}\Gamma'\,,\qquad \vec r\in\Gamma\,.
\label{eq:varrhoint}
\end{multline}
When $\vec J_{\rm s}$ and $\varrho_{\rm s}$ are solutions
to~(\ref{eq:MFIE}) and~(\ref{eq:varrhoint}), then $\vec E(\vec r)$
in~(\ref{eq:Erep}) is a solution to the
system~(\ref{eq:PDE1}),~(\ref{eq:PDE2}), and~(\ref{eq:BC}).

The Fredholm second kind integral equation~(\ref{eq:varrhoint}) and
its exterior counterpart are here denoted the ChIE. Over the last
decade, the ChIE has become a tool for dealing with certain numerical
problems known as ``low-frequency breakdown'' and which occur when the
MFIE, or the related formulations EFIE and CFIE, are used for exterior
electromagnetic scattering at low frequencies~\cite{Vicoetal13}.
Low-frequency breakdown is caused by decoupling of electric and
magnetic fields. It manifests itself when integral equations are
solved or when fields are reconstructed. More generally, the ChIE has
been combined with the MFIE~\cite{HelsKarl15,Vicoetal13}, with the
EFIE~\cite{BremGimb13}, and with the CFIE~\cite{BendColl12,TaskOija06}
for PEC surfaces and with the EFIE and the MFIE for penetrable
objects~\cite{TaskOija06}. The combination of the MFIE and the ChIE
for exterior problems is denoted the ECCIE in~\cite{Vicoetal13}.
Low-frequency breakdown is not an issue when computing eigenfields
since cavity resonances are wave phenomena with a strong coupling
between electric and magnetic fields. Our reasons for preferring the
ChIE over~(\ref{eq:diffj}) have, as in~\cite{HelsKarl15}, to do with
convergence order and achievable accuracy.

In accordance with \cite{HelsKarl15} we split~(\ref{eq:MFIE}) into the
two coupled scalar equations
\begin{equation}
\begin{split}
\left(I+K_1\right)J_{\tau}(\vec r)+{\rm i}K_2J_{\theta}(\vec r)=0\,,\\
{\rm i}K_3J_{\tau}(\vec r)+\left(I+K_4\right)J_{\theta}(\vec r)=0\,.
\end{split}
\label{eq:comp12}
\end{equation}
and write \eqref{eq:varrhoint} as
\begin{equation}
(I-2K_{\nu})\varrho_{\rm s}(\vec r)=
-2{\rm i}kS_5J_{\tau}(\vec r)+2kS_6J_{\theta}(\vec r)\,.
\label{eq:varrhoint2}
\end{equation}
Here $I$ is the identity. The operators $K_\alpha$,
$\alpha=1,2,3,4,\nu$ are of the double-layer type~(\ref{eq:Kdefa})
with~(\ref{eq:Kdefb}). The operators $S_5$ and $S_6$ are of the
single-layer type~(\ref{eq:Sdefa}) with~(\ref{eq:Sdefb}). See
Appendix~\ref{sec:explicitS} for their explicit expressions.

\section{Fourier series expansions}
\label{sec:Fourier}

The aim of this paper is to present a high-order convergent and
accurate discretization scheme to solve the MFIE and to evaluate
electric eigenfields, normalized by~(\ref{eq:Enorm}). We employ a
Fourier--Nyström technique where the first step is an azimuthal
Fourier transformation of the MFIE system~(\ref{eq:comp12}) and
(\ref{eq:varrhoint2}) and of the system for the decomposed electric
field~(\ref{eq:comp13}).

We define the azimuthal Fourier coefficients for 2$\pi$-periodic
functions
\begin{align}
g_n(r)&=\frac{1}{\sqrt{2\pi}}\int_{-\pi}^{\pi}e^{-{\rm i}n\theta}
g({\vec r})\,{\rm d}\theta\,, 
\label{eq:gF}\\
G_n(r,r')&=
\frac{1}{\sqrt{2\pi}}\int_{-\pi}^{\pi}e^{-{\rm i}n(\theta-\theta')}
G({\vec r},{\vec r}')\,{\rm d}(\theta-\theta')\,,
\label{eq:GF}
\end{align}
where $g(\vec r)$ represents functions like $\varrho_{\rm s}(\vec r)$,
$J_\tau(\vec r)$, $J_\theta(\vec r)$, $E_\rho(\vec r)$, $E_\theta(\vec
r)$, and $E_z(\vec r)$ and where $G({\vec r},{\vec r}')$ represents
$K_\alpha({\vec r},{\vec r}')$, $D_\alpha({\vec r},{\vec r}')$,
$S_\alpha({\vec r},{\vec r}')$, $Z_\alpha({\vec r},{\vec r}')$, and
$P(|{\vec r}-{\vec r}'|)$. The subscript $n$ is the azimuthal index,
$n=0,\pm 1,\pm 2,\ldots$.

We also define the modal integral operators $K_{\alpha n}$ and
$S_{\alpha n}$ in terms of their corresponding Fourier coefficients
$K_{\alpha n}(r,r')$ and $S_{\alpha n}(r,r')$ as
\begin{align}
K_{\alpha n}g_n(r)=
\sqrt{2\pi}\int_\gamma K_{\alpha n}(r,r')g_n(r')\rho'\,{\rm d}\gamma'\,,
\label{eq:Kndef}\\
S_{\alpha n}g_n(r)=
\sqrt{2\pi}\int_\gamma S_{\alpha n}(r,r')g_n(r')\rho'\,{\rm d}\gamma'\,.
\label{eq:Sndef}
\end{align}

Expansion and integration of~(\ref{eq:comp12}) over $\theta'$ now give
the system of modal integral equations
\begin{equation}
\begin{split}
\left(I+K_{1n}\right)J_{\tau n}(r)
+{\rm i}K_{2n}J_{\theta n}(r)=0\,, \\
 {\rm i}K_{3n}J_{\tau n}(r)+\left(I+K_{4n}\right)J_{\theta n}(r)=0\,,
\end{split}
\label{eq:comp12F}
\end{equation}
where $r\in\gamma$. Analogously, the modal version of
(\ref{eq:varrhoint2}) is
\begin{equation}
\left(I-2K_{\vec\nu n}\right)\varrho_{{\rm s}n}(r)=
-2{\rm i}kS_{5n}J_{\tau n}(r)+2kS_{6n}J_{\theta n}(r)\,.
\label{eq:varrhoint2F}
\end{equation}
The modal representation of the electric field in (\ref{eq:comp13})
is
\begin{equation}
\begin{split}
E_{\rho n}(r)  &= {\rm i}kS_{7n}J_{\tau n}(r)
              +kS_{8n}J_{\theta n}(r)+K_{11n}\varrho_{{\rm s}n}(r)\,,\\
E_{\theta n}(r)&= kS_{9n}J_{\tau n}(r) + 
 {\rm i}kS_{10n}J_{\theta n}(r)+{\rm i}K_{12n}\varrho_{{\rm s}n}(r)\,,\\
E_{z n}(r)     &= {\rm i}kS_{11n}J_{\tau n}(r)
                        +K_{13n}\varrho_{{\rm s}n}(r)\,,
\end{split}
\label{eq:comp13F}
\end{equation}
where $r\notin\gamma$.

In what follows, we sometimes present only modal expressions for
operators and fields. The reasons being the close resemblance between
expressions in $\mathbb{R}^3$ and modal expressions, visible in the
examples above, and that we seek eigenfields
\begin{equation}
\vec E_n(\vec r)=\left(\vec\rho E_{\rho n}(r)
+\vec\theta E_{\theta n}(r)+\vec zE_{z n}(r)\right)
\frac{e^{{\rm i}n\theta}}{\sqrt{2\pi}}
\label{eq:Eeig}
\end{equation}
for one azimuthal mode at a time.

\section{Conversion of the normalization integral}
\label{sec:normal}

The normalization (volume) integral in (\ref{eq:Enorm}) can be
converted into a sum of line integrals over $\gamma$ that is more
suitable for numerical evaluation. The line integrals contain Fourier
coefficients of an electric scalar potential $\Psi(\vec r)$ and a
magnetic vector potential $\boldsymbol{\it\Lambda}(\vec r)$.
In~\cite[Appendix A]{HelsKarl15}, using results
from~\cite{BarnHass14}, this conversion is done in the context of
magnetic eigenfields. Since for each eigenwavenumber the electric and
magnetic eigenfield energies are equal, the expression
in~\cite{HelsKarl15} applies equally well to electric eigenfields.
With $\vec E_n(\vec r)$ as in~(\ref{eq:Eeig}) the converted expression
reads
\begin{multline}
\|\vec E_n\|^2=
k^2\int_A\left(|\boldsymbol{\it\Lambda}_n|^2-|\Psi_n|^2\right)\rho\,{\rm d}A
+\int_\gamma\left(\boldsymbol{\it\Lambda}_n^\ast\cdot\vec J_{{\rm s}n}
+{\rm i}k\Lambda_{\nu n}^\ast\Psi_n\right)\rho\,{\rm d}\gamma\,,
\label{eq:magne2}
\end{multline}
\newpage\noindent
where
\begin{align}
k^2\int_A|\Psi_n|^2\rho \,{\rm d}A&=
-\frac{1}{2}\int_\gamma\frac{\nu\cdot r}{\rho}\left(n^2-k^2\rho^2\right)
|\Psi_n|^2\,{\rm d}\gamma
\nonumber\\
&\qquad-\frac{1}{2}\int_\gamma\nu\cdot r
 \left(\left|(\partial_{\vec\tau}\Psi)_n\right|^2
      -\left|(\partial_{\vec\nu^+}\Psi)_n\right|^2\right)
\rho\,{\rm d}\gamma
\nonumber\\
&+\frac{1}{2}\int_\gamma{\rm Re}
\left\{\left(2\tau\cdot r(\partial_{\vec\tau}\Psi^\ast)_n+
\Psi^\ast_n\right)(\partial_{\vec\nu^+}\Psi)_n\right\}
\rho\,{\rm d}\gamma\,,
\label{eq:Psq2}\\
k^2\int_A|\boldsymbol{\it\Lambda}_n|^2\rho\,{\rm d}A
&=-\frac{1}{2}\int_\gamma\frac{\nu\cdot r}{\rho}\left(
(n^2-k^2\rho^2)|\boldsymbol{\it\Lambda}_n|^2\right)\,{\rm d}\gamma\nonumber\\
&-\frac{1}{2}\int_\gamma\frac{\nu\cdot r}{\rho}\left(
|\Lambda_{\rho n}|^2+|\Lambda_{\theta n}|^2
-4n\,{\rm Im}\left\{ \Lambda_{\rho n}^\ast\Lambda_{\theta n}\right\}
\right)\,{\rm d}\gamma\nonumber\\
&\qquad-\frac{1}{2}\int_\gamma\nu\cdot r\left(
 \left|(\partial_{\vec\tau}\boldsymbol{\it\Lambda})_n\right|^2
-\left|(\partial_{\vec\nu^+}\boldsymbol{\it\Lambda})_n\right|^2\right)
 \rho\,{\rm d}\gamma\nonumber\\
&+\frac{1}{2}\int_\gamma{\rm Re}\left\{
(2\tau\cdot r(\partial_{\vec\tau}\boldsymbol{\it\Lambda}^\ast)_n
+\boldsymbol{\it\Lambda}_n^\ast)
\cdot(\partial_{\vec\nu^+}\boldsymbol{\it\Lambda})_n
\right\}\rho\,{\rm d}\gamma\,,
\label{eq:Asq2}
\end{align}
and where directional derivatives of a function $g(\vec r)$ are
abbreviated as
\begin{equation}
\begin{split}
\partial_{\vec\tau}g(\vec r)&=
\vec\tau\cdot\nabla g(\vec r)\,,\qquad \vec r\in\Gamma\,,\\
\partial_{\vec\nu^+}g(\vec r^\circ)&=
\lim_{V\ni \vec r\to \vec r^\circ}
\vec\nu^\circ\cdot\nabla g(\vec r)\,,\qquad \vec r^\circ\in\Gamma\,.
\end{split}
\end{equation}

In order to evaluate~(\ref{eq:magne2}) from the solution
to~(\ref{eq:comp12F}) and \eqref{eq:varrhoint2F}, the Fourier
coefficients of $\Psi(\vec r)$ and $\boldsymbol{\it\Lambda}(\vec r)$,
and their derivatives with respect to $\vec\tau$ and $\vec\nu^+$, need
to be related to $\varrho_{{\rm s} n}(r)$, $J_{\tau n}(r)$, and
$J_{\theta n}(r)$. Partial information on these relations, along with
derivations, can be found in~\cite[Appendix B]{HelsKarl15}. We now
give complete information without derivations.

Two different decompositions of $\boldsymbol{\it\Lambda}_n(r)$ are needed
\begin{equation}
\begin{split}
\boldsymbol{\it\Lambda}_n(r)&=\vec\tau\Lambda_{\tau n}(r)
                             +\vec\theta\Lambda_{\theta n}(r)
                             +\vec\nu\Lambda_{\nu n}(r)\,,\\
\boldsymbol{\it\Lambda}_n(r)&=\vec\rho\Lambda_{\rho n}(r)
                             +\vec\theta\Lambda_{\theta n}(r)
                             +\vec z\Lambda_{z n}(r)\,.
\end{split}
\end{equation}
Expressions for $\boldsymbol{\it\Lambda}_n(r)$ are in these bases
\begin{equation}
\begin{split}
\Lambda_{\tau n}(r)&=         S_{1n}J_{\tau n}(r)
                      +{\rm i}S_{2n}J_{\theta n}(r)\,,\\
\Lambda_{\theta n}(r)&={\rm i}S_{3n}J_{\tau n}(r)
                             +S_{4n}J_{\theta n}(r)\,,\\
\Lambda_{\theta 0}(r)&=0\,,\\
\Lambda_{\nu n}(r)&=          S_{5n}J_{\tau n}(r)
                             +{\rm i}S_{6n}J_{\theta n}(r)\,,\\
\Lambda_{\rho n}(r)&=         S_{12n}J_{\tau n}(r)
                      +{\rm i}S_{13n}J_{\theta n}(r)\,,\\
\Lambda_{z n}(r)&=            S_{14n}J_{\tau n}(r)\,.
\end{split}
\label{eq:Lexpr}
\end{equation}
The modal operators $S_{\alpha n}$, $\alpha=1,\ldots,6,12,13,14$, are
defined via~(\ref{eq:Sndef}), (\ref{eq:GF}), (\ref{eq:Sdefb}) and
Appendix~\ref{sec:explicitS}.

Expressions for the Fourier coefficients of the cylindrical components
of the derivatives of $\boldsymbol{\it\Lambda}(\vec r)$ with respect
to $\vec\tau$ and $\vec\nu^+$ are
\begin{equation}
\begin{split}
(\partial_{\vec\tau}\boldsymbol{\it\Lambda})_{\rho n}(r)&=
 K_{14n}J_{\tau n}(r)+{\rm i}K_{15n}J_{\theta n}(r)\,,\\
(\partial_{\vec\tau}\boldsymbol{\it\Lambda})_{\theta n}(r)&=
 {\rm i}K_{16n}J_{\tau n}(r)+K_{17n}J_{\theta n}(r)\,,\\
(\partial_{\vec\tau}\boldsymbol{\it\Lambda})_{z n}(r)&=
K_{18n}J_{\tau n}(r)\,,\\
(\partial_{\vec\nu^+}\boldsymbol{\it\Lambda})_{\rho n}(r)&=
\frac{1}{2} \nu_z J_{\tau n}(r)+K_{19n}J_{\tau n}(r)
+{\rm i}K_{20n}J_{\theta n}(r)\,,\\
(\partial_{\vec\nu^+}\boldsymbol{\it\Lambda})_{\theta n}(r)&=
\frac{1}{2}J_{\theta n}(r)
 +{\rm i}K_{21n}J_{\tau n}(r)+K_{22n}J_{\theta n}(r)\,,\\
(\partial_{\vec\nu^+}\boldsymbol{\it\Lambda})_{z n}(r)&=
-\frac{1}{2} \nu_{\rho}J_{\tau n}(r)+K_{23n}J_{\tau n}(r)\,.
\end{split}
\label{eq:LDexpr}
\end{equation}
The modal operators $K_{\alpha n}$, $\alpha=14,\ldots,23$, are defined
via~(\ref{eq:Kndef}), (\ref{eq:GF}), (\ref{eq:Kdefb}) and
Appendix~\ref{sec:explicitS}.

Expressions for the Fourier coefficients of $\Psi(\vec r)$ and its
derivatives are~\cite{HelsKarl15}
\begin{equation}
\begin{split}
\Psi_n(r)&=\frac{k}{n}\rho\Lambda_{\theta n}(r)\,,\qquad n\ne 0\,,\\
\Psi_0(r)&=S_{\varsigma 0}\varrho_{{\rm s}0}(r)\,,\\
(\partial_{\vec\tau}\Psi)_n(r)&={\rm i}k\Lambda_{\tau n}(r)\,,\\
(\partial_{\vec\nu^+}\Psi)_n(r)&=
\varrho_{{\rm s}n}(r)+{\rm i}k\Lambda_{\nu n}(r)\,,
\end{split}
\label{eq:Pexpr}
\end{equation}
with $S_{\varsigma 0}$ defined via~(\ref{eq:Sndef}), (\ref{eq:GF}),
(\ref{eq:Sdefb}) and Appendix~\ref{sec:explicitS}.

\section{Fourier coefficients of static kernels in analytic form}
\label{sec:azimuthn}

When $r$ and $r'$ are far from each other, all kernels $K_\alpha(\vec
r,\vec r')$ and $S_\alpha(\vec r,\vec r')$ are smooth functions of
$\theta-\theta'$ and we evaluate the corresponding Fourier
coefficients $K_{\alpha n}(r,r')$ and $S_{\alpha n}(r,r')$, needed
in~(\ref{eq:comp12F}), (\ref{eq:varrhoint2F}), (\ref{eq:comp13F}) and
(\ref{eq:magne2}), from~(\ref{eq:GF}) using discrete Fourier transform
techniques (FFT). When $r$ and $r'$ are close, the kernels vary more
rapidly and FFT alone is not efficient. Instead we split each
$K_\alpha(\vec r,\vec r')$ and $S_\alpha(\vec r,\vec r')$ into two
parts: a smooth part, which is transformed via FFT, and a rapidly
varying part, which is transformed by convolution of $D_{\alpha
  n}(r,r')$ and $Z_{\alpha n}(r,r')$ with parts of $P_n(r,r')$.
See~\cite[Section 6]{HelsKarl14} for details on this splitting
and~\cite[Section 12.1]{HelsKarl14} for a definition of when $r$ and
$r'$ are considered close.

The coefficients $D_{\alpha n}(r,r')$ and $Z_{\alpha n}(r,r')$ can be
expressed in terms of half-integer degree Legendre functions of the
second kind~\cite[Equation 8.713.1]{Grad07}
\begin{equation}
\mathfrak{Q}_{n-\frac{1}{2}}(\chi)=
\int_{-\pi}^{\pi}\frac{\cos(nt)\,{\rm d}t}
{\sqrt{8\left(\chi-\cos(t)\right)}}\,.
\label{eq:Qndef}
\end{equation}
We use these analytic expressions in the convolutions where, in our
setting,
\begin{equation}
\chi=1+\frac{|r-r'|^2}{2\rho\rho'}\,.
\label{eq:chidef}
\end{equation}
The functions $\mathfrak{Q}_{n-\frac{1}{2}}(\chi)$, with real
arguments $\chi\ge 1$ and often with $\cosh(\chi)$ substituted for
$\chi$, may also be called toroidal
functions~\cite[Section 8.850]{Grad07}, \cite[page 201]{Magn66}, ring
functions~\cite[Section 8.11]{Abra72}, or toroidal
harmonics~\cite{Gil00}. They are symmetric with respect to $n$,
exhibit logarithmic singularities at $\chi=1$, and are relatively
cheap to evaluate.

The particular combinations of toroidal functions
\begin{equation}
\mathfrak{R}_n(\chi)=\frac{2n-1}{\chi+1}\left(
\chi\mathfrak{Q}_{n-\frac{1}{2}}(\chi)
-\mathfrak{Q}_{n-\frac{3}{2}}(\chi)\right)
\label{eq:Rndef}
\end{equation}
play an important role in our analytic expressions. They multiply
functions which may exhibit Cauchy-type singularities at $\chi=1$. The
$\mathfrak{R}_n(\chi)$ are finite at $\chi=1$, but have logarithmic
singularities in their first (right) derivatives.

The toroidal functions can be evaluated via a recursion whose forward
form is
\begin{equation}
\mathfrak{Q}_{n-\frac{1}{2}}(\chi)=
 \frac{4n-4}{2n-1}\chi\mathfrak{Q}_{n-\frac{3}{2}}(\chi)
-\frac{2n-3}{2n-1}\mathfrak{Q}_{n-\frac{5}{2}}(\chi)\,,\qquad n=2,3,\ldots\,.
\label{eq:forwrec}
\end{equation}
When $1<\chi<1.0005$ we use~(\ref{eq:forwrec}) as it stands. When
$\chi\ge 1.0005$, and for stability reasons, we use a backward form
of~(\ref{eq:forwrec}). A short {\sc Matlab} code which evaluates
$\mathfrak{Q}_{-\frac{1}{2}}(\chi)$ and
$\mathfrak{Q}_{\frac{1}{2}}(\chi)$ is presented in~\cite[Appendix
C]{HelsKarl15}. See Appendix~\ref{sec:explicitF} for a complete
description of how all coefficients $D_{\alpha n}(r,r')$ and
$Z_{\alpha n}(r,r')$, needed in the present work, are related to
$\mathfrak{Q}_{n-\frac{1}{2}}(\chi)$.

\section{An overview of the discretization}
\label{sec:discrete}

Our Fourier--Nyström discretization scheme is very similar to the
scheme used in~\cite{HelsKarl15}. That scheme, in turn, builds on the
schemes developed in~\cite{HelsKarl14,Youn12} in a pure Helmholtz
setting. This section only gives a brief review.

The FFT operations are, basically, controlled by two problem dependent
integers $N$ and $N_{\rm con}$, with $N\ge N_{\rm con}$, as follows:
we use $2N+1$ equispaced points in the azimuthal Fourier transforms of
kernels when $r$ and $r'$ are far from each other, we use $2N_{\rm
  con}+1-n$ terms in the truncated convolutions~\cite[Equation
(27)]{HelsKarl14}, and we use $2N_{\rm con}+1$ equispaced points in
the azimuthal Fourier transforms of smooth parts of kernels when $r$
and $r'$ are close. The value of $N$ is chosen in an ad hoc manner.
The value of $N_{\rm con}$ is determined by the decay of the Fourier
coefficients of the test function
\begin{equation}
g(\theta)=\frac{\sin(2k\rho_{\rm max}\sin(\theta/2))}{\sin(\theta/2)}\,,
\label{eq:gtest}
\end{equation}
where $\rho_{\rm max}$ is the largest value of $\rho$ on $\gamma$, so
that $N_{\rm con}$ is the number of Fourier coefficients $g_n$,
$n=1,2,\ldots$, with $g_n/g_0>10\epsilon_{\rm mach}$. When $r$ and
$r'$ are in the vicinity of corners, however, we may use smaller
values of $N_{\rm con}$. See, further, Section~\ref{sec:particul}. We
note, but do not generally exploit, that more elaborate adaptivity in
the control of the FFT operations can lead to substantial
computational savings.

Our Nyström discretization of~(\ref{eq:comp12F}),
(\ref{eq:varrhoint2F}), (\ref{eq:comp13F}) and (\ref{eq:magne2})
relies on an underlying panel-based 16-point Gauss--Legendre
quadrature with a mesh of $n_{\rm pan}$ quadrature panels on $\gamma$.
The $16n_{\rm pan}$ discretization points play the role of both target
points $r_i$ and source points $r_j$. The underlying quadrature is
used in a conventional way when $r_i$ and $r_j$ are far from each
other. When $r_i$ and $r_j$ are close and convolution is used, see
Section~\ref{sec:azimuthn}, an explicit kernel-split special
quadrature is activated. Analytical information about the
singularities in $K_{\alpha n}(r,r')$ and $S_{\alpha n}(r,r')$ is
exploited in the construction of 16th order accurate weight
corrections, computed on the fly. As to some extent compensate for the
loss of convergence order that comes with the special quadrature, a
procedure of temporary mesh refinement (upsampling) is adopted.
See~\cite{HelsHols14} for additional information on quadrature
construction and upsampling.

It is worth emphasizing that all $K_{\alpha n}(r,r')$ and $S_{\alpha
  n}(r,r')$ contain some sort of singularities at $r=r'$ and that
these singularities are inherited by the corresponding $D_{\alpha
  n}(r,r')$ and $Z_{\alpha n}(r,r')$ listed in
Appendix~\ref{sec:explicitF}. The singularities are generally of
logarithmic type, native to $\mathfrak{Q}_{n-\frac{1}{2}}(\chi)$, with
the exceptions that the coefficients in Appendix~\ref{sec:explicitF}
that contain the function
\begin{equation}
d(v)=\frac{v\cdot(r-r')}{|r-r'|^2}\,,\qquad v=\tau,\nu,\hat{\rho},\hat{z}\,,
\end{equation}
may exhibit Cauchy-type singularities as $r$ approaches $r'$ and those
coefficients that are proportional to $\mathfrak{R}_n(\chi)$ have
logarithmic-type singularities only in their first derivatives. The
quadratures constructed in~\cite{HelsHols14,HelsKarl14} cover all
these situations.

\section{Recursively compressed inverse preconditioning}
\label{sec:RCIP}

Spectral properties of integral operators in boundary integral
equations are often sensitive to boundary smoothness. The very nature
of solutions may be affected by a change in smoothness as may the
performance of numerical solvers. For example, the introduction of
boundary singularities such as edges and corners can induce diverging
asymptotics in layer densities. Intense and costly mesh refinement is
then needed for resolution, which may lead to the loss of stability.
See~\cite{Gill14} for a review of recently developed numerical
techniques to deal with this problem.

RCIP is one of the techniques discussed in~\cite{Gill14}. It can be
viewed as a general method to enhance the performance of panel-based
Nyström discretization schemes. Roughly speaking, for Fredholm second
kind integral equations, RCIP obtains solutions on piecewise smooth
curves with the same accuracy and at about the same cost as with which
solutions normally are obtained on smooth curves. The RCIP method
originated in 2008 in the context of solving Laplace's equation in
piecewise smooth planar domains~\cite{Hels08} and has since then been
improved and extended as to apply to a variety of boundary value
problems.

A comprehensive description of the RCIP method is given in the
tutorial~\cite{Hels13}. This section first gives a brief summary and
then focuses on some details particular to the MFIE
system~(\ref{eq:comp12F}) and (\ref{eq:varrhoint2F}) and to the
converted normalization integral~(\ref{eq:magne2})
with~(\ref{eq:Psq2}) and (\ref{eq:Asq2}).

\subsection{Basics of the RCIP method}
\label{sec:basics}

Assume the following: we have an integral representation of a field
$U(r)$, $r\in\mathbb{R}^2\setminus\gamma$, in terms of an unknown
layer density $\sigma(r)$ on a piecewise smooth boundary $\gamma$. On
$\gamma$ there are a number $n_{\rm crn}$ of corners with vertices
$\gamma_j$, $j=1,\ldots,n_{\rm crn}$. The integral representation
together with boundary conditions lead to a Fredholm second kind
integral equation
\begin{equation}
\left(I+K\right)\sigma(r)=g(r)\,,\qquad r\in\gamma\,,
\label{eq:inteq1}
\end{equation}
where $K$ is an integral operator with kernel $K(r,r')$ on $\gamma$.
The operator $K$ is compact away from the corners. The function $g(r)$
is a right hand side with the same smoothness properties as $\gamma$.
We also assume that there is a relatively coarse mesh with $n_{\rm
  pan}$ coarse quadrature panels of approximately equal length
constructed on $\gamma$. The purpose of the coarse mesh is to allow
for a discretization that resolves $g(r)$, $\gamma$, and $K(r,r')$ for
$r$ far away from $r'$.

We split the kernel
\begin{equation}
K(r,r')=K^\star(r,r')+K^\circ(r,r')
\label{eq:ksplit}
\end{equation}
in such a way that $K^\star(r,r')$ is zero except for when $r$ and
$r'$ both lie within a distance of two coarse quadrature panels from
the same $\gamma_j$. In this latter case $K^{\circ}(r,r')$ is zero.
The kernel splitting~(\ref{eq:ksplit}) corresponds to an operator
splitting
\begin{equation}
K=K^\star+K^\circ\,,
\label{eq:osplit}
\end{equation}
where $K^{\circ}$ is a compact operator. The variable substitution
\begin{equation}
\sigma(r)=\left(I+K^\star\right)^{-1}\tilde{\sigma}(r)
\label{eq:subst}
\end{equation}
lets us rewrite~(\ref{eq:inteq1}) as a right preconditioned integral
equation
\begin{equation}
\tilde{\sigma}(r)+K^\circ(I+K^\star)^{-1}\tilde{\sigma}(r)
= g(r)\,,\qquad r\in \gamma\,,
\label{eq:inteq2}
\end{equation}
where the operator composition $K^\circ(I+K^\star)^{-1}$ is compact.

The functions $\tilde{\sigma}(r)$ and $g(r)$ and the operator
$K^{\circ}$ in~(\ref{eq:inteq2}) should be easy to discretize and to
resolve on the coarse mesh. Only the inverse $(I+K^\star)^{-1}$ needs
a fine mesh for its resolution. This fine mesh is constructed from the
coarse mesh by, for each vertex $\gamma_j$, choosing a number $n_{{\rm
    sub}j}$ and letting the panels closest to $\gamma_j$ be $n_{{\rm
    sub}j}$ times repeatedly subdivided. The size of $n_{{\rm sub}j}$
is determined by the behavior of $\sigma(r)$ close to $\gamma_j$ and
by application-specific needs for resolution. The discretization
of~(\ref{eq:inteq2}) can then be carried out as
\begin{equation}
\left({\bf I}_{\rm coa}+{\bf K}_{\rm coa}^{\circ}{\bf R}\right)
\tilde{\boldsymbol{\sigma}}_{\rm coa}={\bf g}_{\rm coa}\,,
\label{eq:inteq3}
\end{equation}
where the compressed weighted inverse matrix ${\bf R}$ is given by
\begin{equation}
{\bf R}=
{\bf P}_W^T\left({\bf I}+{\bf K}^\star\right)_{\rm fin}^{-1}
{\bf P}\,.
\label{eq:R}
\end{equation}
In~(\ref{eq:inteq3}) and~(\ref{eq:R}) the subscript ``coa'' indicates
a grid on the coarse mesh, the subscript ``fin'' indicates a grid on
the fine mesh, the prolongation matrix ${\bf P}$ performs polynomial
interpolation from the coarse grid to the fine grid and ${\bf P}_W^T$
is the transpose of a weighted prolongation matrix such that
\begin{equation}
{\bf P}_W^T{\bf P}={\bf I}_{\rm coa}\,.
\end{equation}
See~\cite[Sections 4 and 5]{Hels13} for details. With 16-point
composite quadrature the system size in~(\ref{eq:inteq3}) is $16n_{\rm
  pan}\times 16n_{\rm pan}$. The matrix ${\bf R}$ differs from the
identity matrix by having $n_{\rm crn}$ diagonal blocks ${\bf
  R}^{(j)}$, $j=1,\ldots,n_{\rm crn}$, of size $64\times 64$.

Once~(\ref{eq:inteq3}) is solved for $\tilde{\boldsymbol{\sigma}}_{\rm
  coa}$, a discrete weight-corrected version of the original layer
density is obtained from
\begin{equation}
\hat{\boldsymbol{\sigma}}_{\rm coa}={\bf R}
\tilde{\boldsymbol{\sigma}}_{\rm coa}\,.
\label{eq:wcorr}
\end{equation}
The density $\hat{\boldsymbol{\sigma}}_{\rm coa}$, together with the
composite quadrature, can be used for the accurate discretization of
any integral on $\gamma$ involving $\sigma(r)$ and piecewise smooth
functions. Furthermore, the field $U(r)$ can now be recovered in those
parts of the computational domain that lie away from the vertices
$\gamma_j$ using $\hat{\boldsymbol{\sigma}}_{\rm coa}$ together with
the quadratures of~\cite{HelsHols14,HelsKarl14} in a discretization of
the integral representation for $U(r)$.

Note that in~(\ref{eq:inteq3}), the need for resolution in corners is
not visible. The transformed layer density
$\tilde{\boldsymbol{\sigma}}_{\rm coa}$ should be as easy to solve for
as the original layer density in a discretization of~(\ref{eq:inteq1})
on a smooth $\gamma$. All computational difficulties are gathered in
the construction of the matrix ${\bf R}$.

There are $64+32n_{{\rm sub}j}$ discretization points on the fine grid
within a distance of two coarse panels from the vertex $\gamma_j$.
Judging from an inspection of~(\ref{eq:R}) it seems as if computing
the matrix block ${\bf R}^{(j)}$ should be an expensive and also
unstable undertaking for large $n_{{\rm sub}j}$. Fortunately, ${\bf
  R}^{(j)}$ can be computed via a fast and stable recursion which
relies on hierarchies of small local nested grids around $\gamma_j$
and produces hierarchies of matrices ${\bf R}^{(j)}_i$,
$i=1,\ldots,n_{{\rm sub}j}$, where the last matrix is equal to ${\bf
  R}^{(j)}$. This fast recursion enables the computation of ${\bf
  R}^{(j)}$ at a cost only proportional to $n_{{\rm sub}j}$. This is
the power of the RCIP method.

The fast recursion for ${\bf R}$ can also be run backwards, acting on
$\tilde{\boldsymbol{\sigma}}_{\rm coa}$, for the purpose of
reconstructing the solution $\boldsymbol{\sigma}_{\rm fin}$ to a
straight-forward discretization of~(\ref{eq:inteq1}) on the fine mesh
\begin{equation}
\left({\bf I}_{\rm fin}+{\bf K}_{\rm fin}\right)
\boldsymbol{\sigma}_{\rm fin}={\bf g}_{\rm fin}\,.
\end{equation}
By this, one sees that the information contained in
$\tilde{\boldsymbol{\sigma}}_{\rm coa}$, together with the ${\bf
  R}^{(j)}_i$, is the same as the information contained in
$\boldsymbol{\sigma}_{\rm fin}$. A partial reconstruction of
$\boldsymbol{\sigma}_{\rm fin}$ is needed when $U(r)$ is to be
evaluated close to the vertices $\gamma_j$. See~\cite[Section
9]{Hels13} for a description of the reconstruction procedure.

\subsection{Details particular to the MFIE system}
\label{sec:particul}

The MFIE system~(\ref{eq:comp12F}) and (\ref{eq:varrhoint2F}), which
on block operator form reads
\begin{equation}
\left[ 
\begin{array}{ccc}
I-2K_{\vec\nu n} & 2{\rm i}kS_{5n} & -2kS_{6n}  \\
               0 & I+K_{1n}        & {\rm i}K_{2n} \\
               0 & {\rm i}K_{3n}   & I+K_{4n}
\end{array} 
\right]
\left[ 
\begin{array}{c}
\varrho_{{\rm s}n}(r)\\
J_{\tau n}(r)\\
J_{\theta n}(r)
\end{array} 
\right]
=\left[
\begin{array}{c}
 0\\
 0\\
 0
\end{array} 
\right]\,,
\label{eq:MFIEsys}
\end{equation}
has a more complicated appearance than the model
equation~(\ref{eq:inteq1}). The operator corresponding to $K^\star$
in~(\ref{eq:osplit}), upon discretization, no longer yields a block
diagonal matrix but a sparse block matrix where each vertex $\gamma_j$
generates seven non-zero $64\times 64$ blocks. In practice, this poses
no problems for RCIP. Equation~(\ref{eq:inteq3}) still holds with
\begin{equation}
\begin{split}
&{\bf g}_{\rm coa}={\bf 0}\,,\\
\tilde{\boldsymbol{\sigma}}_{\rm coa}=
\left[
\begin{array}{c}
\tilde{\boldsymbol{\varrho}}_{{\rm s}n}\\
\tilde{\bf J}_{\tau n}\\
\tilde{\bf J}_{\theta n}
\end{array}
\right]_{\rm coa},
&\qquad
{\bf K}^\circ_{\rm coa}=\left[ 
\begin{array}{ccc}
-2{\bf K}^\circ_{\vec\nu n} & 2{\rm i}k{\bf S}^\circ_{5n} & 
-2k{\bf S}^\circ_{6n}  \\
 {\bf 0} & {\bf K}^\circ_{1n}  & {\rm i}{\bf K}^\circ_{2n} \\
 {\bf 0} & {\rm i}{\bf K}^\circ_{3n}   & {\bf K}^\circ_{4n}
\end{array} 
\right]_{\rm coa}.
\end{split}
\label{eq:K3}
\end{equation}
The compressed weighted inverse matrix of~(\ref{eq:R}) is
\begin{multline}
{\bf R}=
\left[ 
\begin{array}{ccc}
{\bf P}_W & {\bf 0} & {\bf 0} \\
{\bf 0} & {\bf P}_W & {\bf 0} \\
{\bf 0} & {\bf 0} & {\bf P}_W
\end{array} 
\right]^T\\
\left[ 
\begin{array}{ccc}
{\bf I}-2{\bf K}^\star_{\vec\nu n} & 2{\rm i}k{\bf S}^\star_{5n} & 
-2k{\bf S}^\star_{6n}  \\
 {\bf 0} & {\bf I}+{\bf K}^\star_{1n}        & {\rm i}{\bf K}^\star_{2n} \\
 {\bf 0} & {\rm i}{\bf K}^\star_{3n}   & {\bf I}+{\bf K}^\star_{4n}
\end{array} 
\right]^{-1}_{\rm fin}
\left[ 
\begin{array}{ccc}
{\bf P} & {\bf 0} & {\bf 0} \\
{\bf 0} & {\bf P} & {\bf 0} \\
{\bf 0} & {\bf 0} & {\bf P}
\end{array} 
\right]
\label{eq:R3}
\end{multline}
and has size $48n_{\rm pan}\times 48n_{\rm pan}$. It can be permuted
as to differ from the identity matrix by $n_{\rm crn}$ diagonal blocks
${\bf R}^{(j)}$ with $7\times 64\times 64$ non-zero entries each. When
computing the ${\bf R}^{(j)}$ via the fast recursion we take advantage
of the sparsity structure in~(\ref{eq:R3}). We also allow for integers
$N_{{\rm con}j}$, controlling FFT operations and convolutions close to
the vertices $\gamma_j$, that may be smaller than the $N_{\rm con}$
used on the coarse grid. These $N_{{\rm con}j}$ are determined as in
Section~\ref{sec:discrete}, but with $\rho_{\rm max}$
of~(\ref{eq:gtest}) replaced with the largest value of $\rho$ on
$\gamma$ within a distance of two coarse panels from $\gamma_j$.

\subsection{Resolving the normalization integral}

The RCIP method provides a tool for the fast and accurate solution of
the MFIE eigensystem~(\ref{eq:MFIEsys}) within the framework of our
Fourier--Nyström scheme. The discretized equation~(\ref{eq:inteq3})
with~(\ref{eq:K3}) and (\ref{eq:R3}) can be used both to find
eigenwavenumbers and to find the corresponding discrete transformed
eigenvectors, that is, non-trivial solutions. The eigenvectors can, in
turn and together with the matrices ${\bf R}^{(j)}_i$, be used to
reconstruct the discrete densities $\boldsymbol{\varrho}_{{\rm s}n}$,
${\bf J}_{\tau n}$, and ${\bf J}_{\theta n}$ on the fine grid. This is
enough to allow for the accurate evaluation of non-normalized electric
eigenfields in the entire computational domain, but not enough to
allow for the accurate evaluation of normalized eigenfields.

The converted normalization integral~(\ref{eq:magne2})
with~(\ref{eq:Psq2}) and (\ref{eq:Asq2}) requires that the Fourier
coefficients of $\Psi(\vec r)$ and $\boldsymbol{\it\Lambda}(\vec r)$,
and their derivatives with respect to $\vec\tau$ and $\vec\nu^+$, are
sufficiently resolved on the fine grid so that their various inner
products and squared moduli can be accurately integrated along
$\gamma$. For a prescribed overall accuracy and for densities
$\varrho_{{\rm s}n}(r)$ or $J_{{\theta n}}(r)$ that diverge at corner
vertices, this poses tougher requirements on panel refinement than
merely demanding that the densities are sufficiently resolved as to be
accurately integrated against piecewise smooth functions. Mappings
from partially reconstructed values of $\boldsymbol{\varrho}_{{\rm
    s}n}$, ${\bf J}_{\tau n}$, and ${\bf J}_{\theta n}$ on the fine
grid to values of these sought Fourier coefficients on the fine grid
can be performed via hierarchical matrix-vector multiplications
similar to, but simpler than, those used for the reconstruction of
$\boldsymbol{\varrho}_{{\rm s}n}$, ${\bf J}_{\tau n}$, and ${\bf
  J}_{\theta n}$ themselves. This procedure uses hierarchies of small
matrices corresponding to the evaluation of all modal operators
present in~(\ref{eq:Lexpr}), (\ref{eq:LDexpr}) and (\ref{eq:Pexpr}) on
the small local nested grids mentioned in Section~\ref{sec:basics}.

\section{Physical fields and edge singularities}
\label{sec:fysikalisk}

This section relates complex valued electric eigenfields in
$\mathbb{R}^3$ of the form~(\ref{eq:Eeig}) to the physical time-domain
standing wave fields that are excited and measured in real life
experiments. To facilitate the interpretation of, so called, field
maps we also review the leading order asymptotic behavior of electric
fields and surface charge and current densities close to edges.

\subsection{Physical time-domain fields.}
\label{sec:phystime}

Every eigenvalue $k^2$ of the system~(\ref{eq:PDE1}),~(\ref{eq:PDE2}),
and~(\ref{eq:BC}), not belonging to the mode $n=0$, is degenerate and
typically corresponds to a two-dimensional subspace of electric
eigenfields. Such an eigenspace can be spanned by two orthonormal
eigenfields $\vec E_n(\vec r)$ and $\vec E_{(-n)}(\vec r)$ of the
form~(\ref{eq:Eeig}), constructed via~(\ref{eq:comp13F}) from the
Fourier coefficients $\varrho_{{\rm s}n}(r)$, $J_{\tau n}(r)$,
$J_{\theta n}(r)$ and $\varrho_{{\rm s}(-n)}(r)$, $J_{\tau(-n)}(r)$,
$J_{\theta(-n)}(r)$ that are non-trivial solutions to
(\ref{eq:comp12F}) and (\ref{eq:varrhoint2F}) at eigenwavenumber $k$
and normalized with~(\ref{eq:Enorm}).

The normalized Fourier coefficients are, in turn, unique only up to a
constant factor of modulus one. In our implementation we choose these
factors such that $J_{\tau n}(r)$ is real and even in $n$, that is,
$J_{\tau(-n)}(r)=J_{\tau n}(r)$. Equations (\ref{eq:comp12F}),
(\ref{eq:varrhoint2F}), (\ref{eq:comp13F}) and the formulas in
Appendix~\ref{sec:explicitF} then imply the following: $J_{\theta
  n}(r)$ is imaginary and odd in $n$; $E_{\rho n}(r)$, $E_{z n}(r)$,
and $\varrho_{{\rm s}n}(r)$ are imaginary and even in $n$; and
$E_{\theta n}(r)$ is real and odd in $n$.

Complex valued standing waves are formed by linear combinations of
Fourier coefficients as
\begin{equation}
\begin{split}
J_{\tau n}^{(\rm e)}(\vec r)&=\frac{a_n}{2}
(J_{\tau n}(r)e^{{\rm i}n\theta}+J_{\tau(-n)}(r)e^{-{\rm i}n\theta}) 
=a_nJ_{\tau n}(r)\cos n\theta\,, \\
J_{\tau n}^{(\rm o)}(\vec r)&=-{\rm i}\frac{a_n}{2}
(J_{\tau n}(r)e^{{\rm i}n\theta}-J_{\tau(-n)}(r)e^{-{\rm i}n\theta})
=a_nJ_{\tau n}(r)\sin n\theta\,, 
\end{split}
\label{eq:JcR3}
\end{equation}
where
\begin{displaymath}
a_0=1/\sqrt{2\pi}\,,\qquad a_n=1/\sqrt{\pi}\,,\qquad n=1,2,\ldots\,,
\end{displaymath}
and superscript $(\rm e)$ and $(\rm o)$ denote ``even'' and ``odd''.
The prefactor $-{\rm i}$ in the second equation of~(\ref{eq:JcR3}) is
to make both surface currents real.

The physical time-domain currents in the $\vec\tau$-direction are now
obtained from
\begin{equation}
\label{eq:timedomainJ}
\begin{split}
J_{\tau n}^{(\rm e)}(\vec r,t)&=
{\rm Re}\{J_{\tau n}^{(\rm e)}(\vec r)e^{-{\rm i}\omega t}\}=
a_nJ_{\tau n}(r)\cos n\theta\cos\omega t\,,\\
J_{\tau n}^{(\rm o)}(\vec r,t)&=
{\rm Re}\{J_{\tau n}^{(\rm o)}(\vec r)e^{-{\rm i}\omega t}\}
=a_nJ_{\tau n}(r)\sin n\theta\cos\omega t\,.
\end{split}
\end{equation}
Expressions for the other physical time-domain quantities are uniquely
composed in the same manner as
\begin{equation}
\begin{split}
\varrho_{{\rm s}n}^{(\rm e)}(\vec r,t)&=
-{\rm i}a_n\varrho_{{\rm s}n}(r)\cos n\theta\sin \omega t\,,\\
J_{\theta n}^{(\rm e)}(\vec r,t)&=
{\rm i}a_nJ_{\theta n}(r)\sin n\theta\cos \omega t\,,\\
E_{\rho n}^{(\rm e)}(\vec r,t)&=
-{\rm i}a_nE_{\rho n}(r)\cos n\theta\sin\omega t\,,\\
E_{\theta n}^{(\rm e)}(\vec r,t)&=
a_nE_{\theta n}(r)\sin n\theta\sin\omega t\,,\\
E_{z n}^{(\rm e)}(\vec r,t)&=
-{\rm i}a_nE_{z n}(r)\cos n\theta\sin\omega t\,,
\end{split}
\label{eq:timedomainE}
\end{equation}
and 
\begin{equation}
\begin{split}
\varrho_{{\rm s}n}^{(\rm o)}(\vec r,t)&=
-{\rm i}a_n\varrho_{{\rm s}n}(r)\sin n\theta\sin\omega t\,,\\
J_{\theta n}^{(\rm o)}(\vec r,t)&=
-{\rm i}a_nJ_{\theta n}(r)\cos n\theta\cos\omega t\,,\\
E_{\rho n}^{(\rm o)}(\vec r,t)&=
-{\rm i}a_nE_{\rho n}(r)\sin n\theta\sin\omega t\,,\\
E_{\theta n}^{(\rm o)}(\vec r,t)&=
-a_nE_{\theta n}(r)\cos n\theta\sin\omega t\,,\\
E_{z n}^{(\rm o)}(\vec r,t)&=
-{\rm i}a_nE_{z n}(r)\sin n\theta\sin\omega t\,.
\end{split}
\label{eq:timedomainO}
\end{equation}
All components of the physical electric eigenfield and its surface
charge density are in phase with respect to time, but 90 degrees out
of phase with the surface current density. In the numerical examples
of Section~\ref{sec:numerical} we present field maps in the $xz$-plane
($\theta=0,\pi$) of the imaginary part of $E_{\rho n}(r)e^{{\rm
    i}n\theta}$ and $E_{z n}(r)e^{{\rm i}n\theta}$, and the real part
of $E_{\theta n}(r)e^{{\rm i}n\theta}$.

\subsection{Asymptotic behavior at edges}
\label{sec:asymp}

Let a corner with vertex $\gamma_j$ have an (inner) opening angle of
$\alpha_j$. Let $\xi_{\rm t}$ be the tangential distance from a point
$r\in\gamma$ to $\gamma_j$ and let $\xi$ be the Euclidean distance
from a point $r\in A\cup\Gamma$ to $\gamma_j$. For $n>0$, we have the
general leading behaviors close to $\gamma_j$
\begin{equation}
\begin{split}
\varrho_{{\rm s}n}(r)\sim \xi_{\rm t}^{p_j-1}\,,\qquad
J_{\tau n}(r)\sim 1\,,\qquad
J_{\theta n}(r)\sim \xi_{\rm t}^{p_j-1}\,,\\
E_{\rho n}(r)\sim \xi^{p_j-1}\,,\qquad
E_{\theta n}(r)\sim 1\,,\qquad
E_{z n}(r)\sim \xi^{p_j-1}\,,
\end{split}
\label{eq:asymp}
\end{equation}
where
\begin{equation}
p_j=\dfrac{\pi}{2\pi-\alpha_j}\,.
\end{equation}
See~\cite{Bibbyetal08} and references therein. The asymptotics of the
Fourier coefficients of $\Psi(\vec r)$ and
$\boldsymbol{\it\Lambda}(\vec r)$, and their derivatives with respect
to $\vec\tau$ and $\vec\nu^+$ are more complicated.

For the special case of $n=0$ one can show that~(\ref{eq:asymp}) holds
with some sharpening: the even fields of (\ref{eq:timedomainE}) have
$J_{\theta 0}(r)\equiv 0$ and $E_{\theta 0}(r)\equiv 0$; the odd
fields of~(\ref{eq:timedomainO}) have $\varrho_{{\rm s}0}(r)\equiv 0$,
$J_{\tau 0}(r)\equiv 0$, and $E_{\rho 0}(r)=E_{z 0}(r)\equiv 0$.

\section{Numerical examples}
\label{sec:numerical}

Our Fourier--Nyström scheme for~(\ref{eq:comp12F}),
(\ref{eq:varrhoint2F}), and (\ref{eq:comp13F}) is implemented in {\sc
  Matlab} and executed on a workstation equipped with an Intel Core
i7-3930K CPU and 64 GB of memory. The weight corrected densities
$\tilde{\boldsymbol{\varrho}}_{{\rm s}n}$, $\tilde{\bf J}_{\tau n}$,
and $\tilde{\bf J}_{\theta n}$ on the coarse grid are obtained
from~(\ref{eq:inteq3}) with~(\ref{eq:K3}) and~(\ref{eq:R3}).  The
densities $\boldsymbol{\varrho}_{{\rm s}n}$, ${\bf J}_{\tau n}$, and
${\bf J}_{\theta n}$ on the fine grid are obtained with
reconstruction~\cite[Section 9]{Hels13}. To enforce~(\ref{eq:Enorm})
we normalize the densities with the value of $\|\vec E_n\|$ obtained
from their insertion in a discretized version of~(\ref{eq:magne2}).

The {\sc Matlab} implementation is standard and relies on built-in
functions. No particular attempts are made at optimizing the code for
speed, except for the use of a few {\tt parfor}-loops (which execute
in parallel). Great care has gone into obtaining intermediate
quantities to high accuracy and to resolve modal integral operators in
corners. The {\it solution time} quoted in the examples below refers
to wall-clock time from when an eigenwavenumber is known and until the
normalized densities $\boldsymbol{\varrho}_{{\rm s}n}$, ${\bf J}_{\tau
  n}$, and ${\bf J}_{\theta n}$ are obtained on the fine grid.

\subsection{Search for eigenwavenumbers}

Eigenwavenumbers are determined using a separate, slimmed down, code
that is cleared from matrices and panel refinement only needed for the
normalization. In what follows we let $k_{n,j}$ be the $j$th smallest
eigenwavenumber for azimuthal index $n$. Our search algorithm for
$k_{n,j}$, with $n$ fixed, is a modification of the ``standard
published method'' described in~\cite[Appendix B]{BarnHass14}. The
standard method is to search along the $k$-axis for (near) zeros of
the lowest singular value $s(k)$ of an appropriate discretized system
matrix ${\bf B}(k)$. Successive parabolic interpolation, which has
convergence rate $r_{\rm c}\approx 1.324$, is applied to $s^2(k)$ and
is safeguarded by the empirical observation that the slope of $s(k)$
appears to have a domain dependent upper bound $C_s$ of size $O(1)$.
In our setting, ${\bf B}(k)$ can be taken as the lower right $32n_{\rm
  pan}\times 32n_{\rm pan}$ part of ${\bf R}^{-1}+{\bf K}_{\rm
  coa}^{\circ}$ with ${\bf R}$ and ${\bf K}_{\rm coa}^{\circ}$ from
(\ref{eq:R3}) and (\ref{eq:K3}).

We modify the standard method, described above, by searching for zeros
of the smallest eigenvalue $\lambda(k)$ of ${\bf B}(k)$ rather than
for (near) zeros of the smallest singular value. We then replace
successive parabolic interpolation applied to $s^2(k)$ with the secant
method applied to $|\lambda(k)|\sgn(\arg(\lambda(k)))$. The slope of
the function $|\lambda(k)|\sgn(\arg(\lambda(k)))$ also appears to have
a domain dependent bound of size $O(1)$, which we denote $C_\lambda$
and use for safeguarding. The secant method has convergence rate
$r_{\rm c}\approx 1.618$. When $C_\lambda$ is chosen correctly and for
a fixed $n$, our modified search algorithm finds all $k_{n,j}$ in a
given $k$-interval typically needing between four and eight iterations
per eigenwavenumber found.

\begin{figure}[!t]
\centering
\includegraphics[height=49mm]{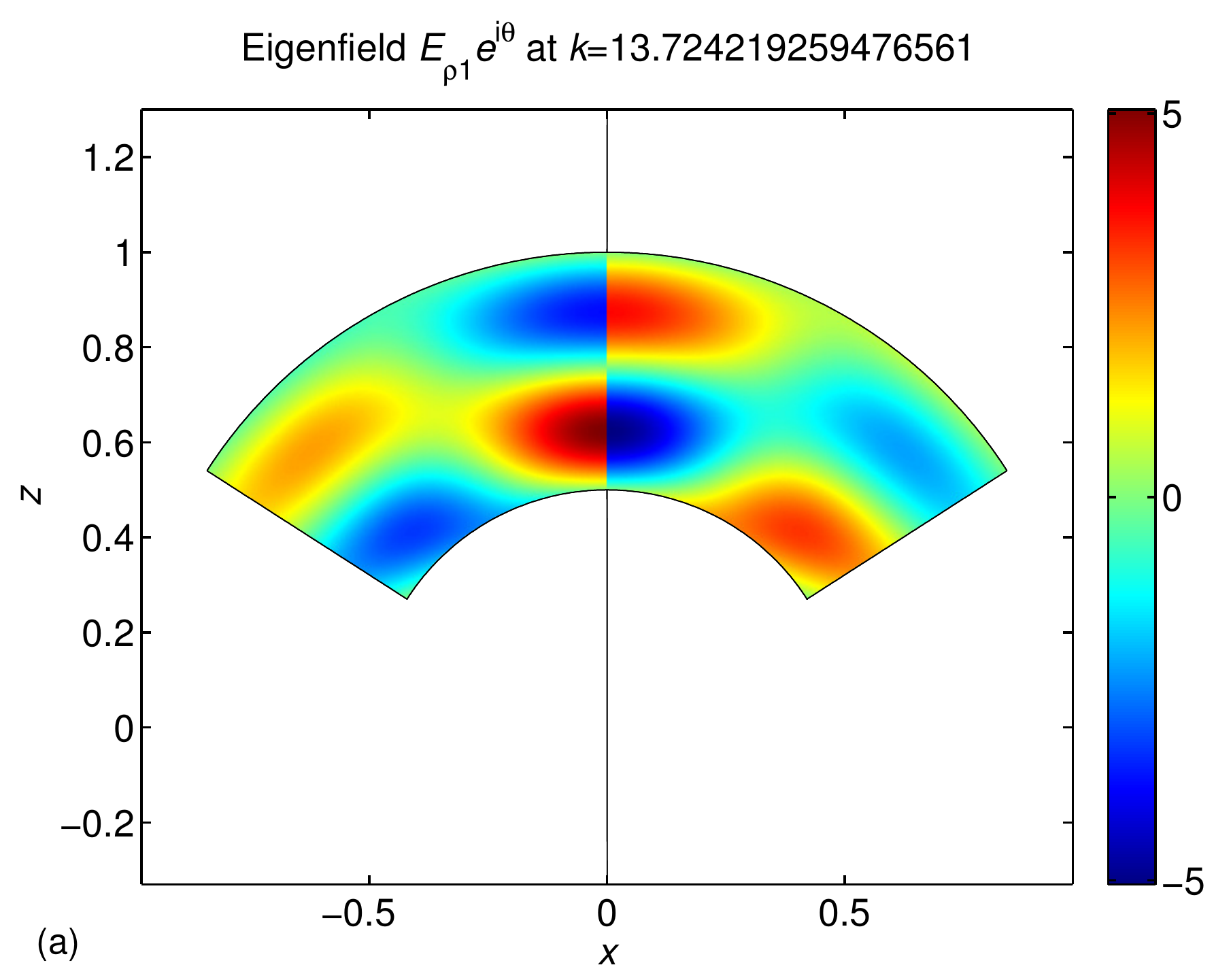}
\includegraphics[height=49mm]{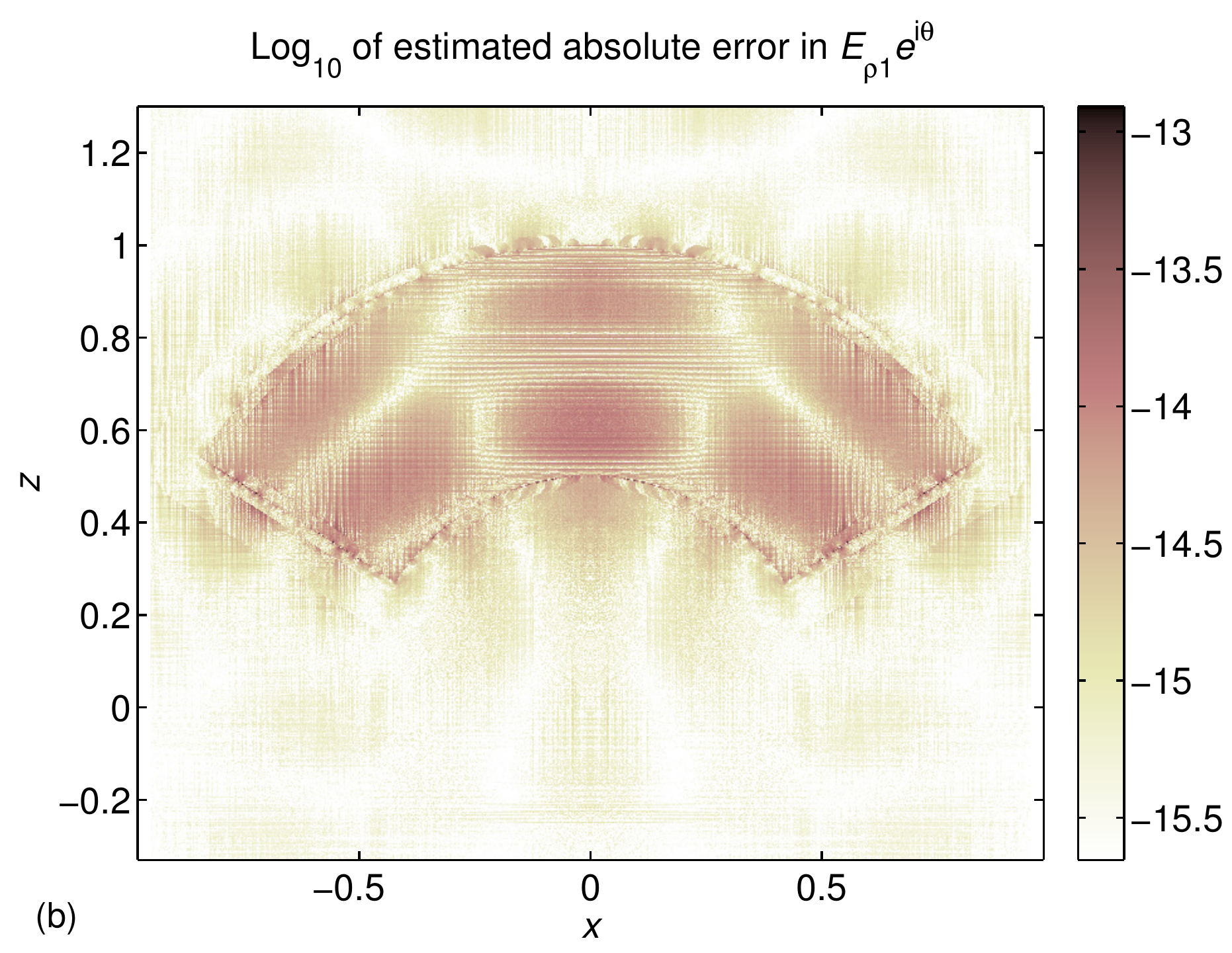}
\includegraphics[height=49mm]{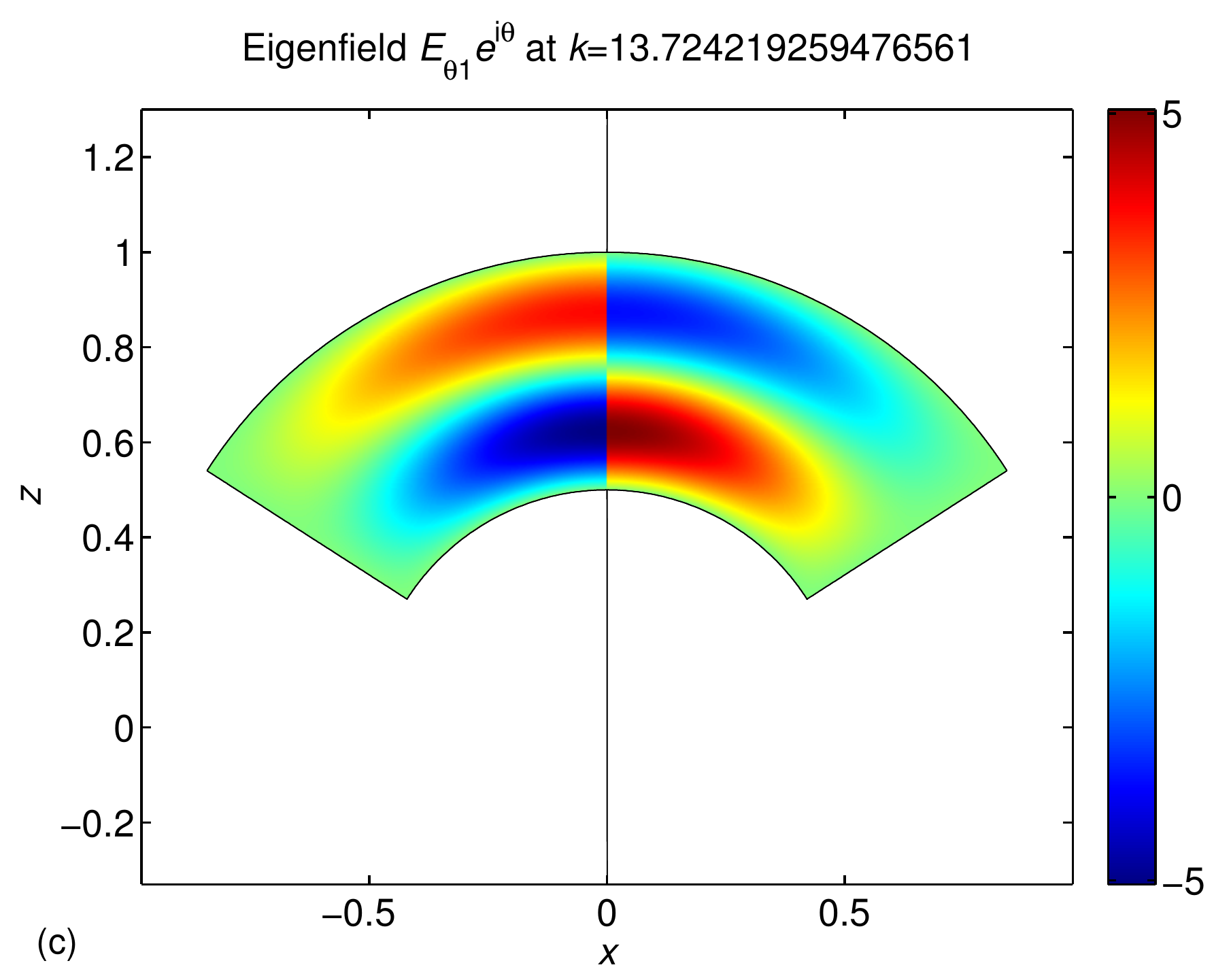}
\includegraphics[height=49mm]{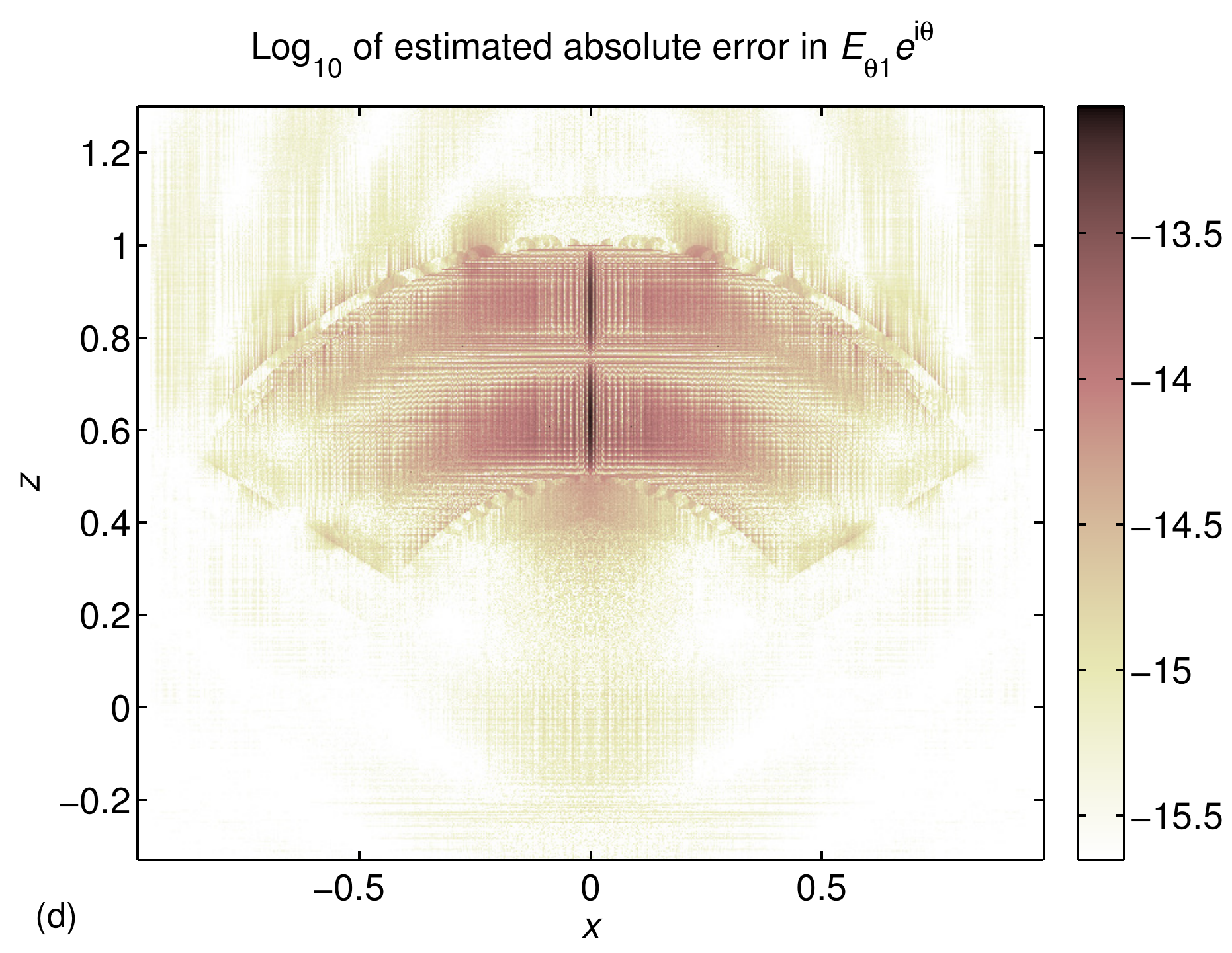}
\includegraphics[height=49mm]{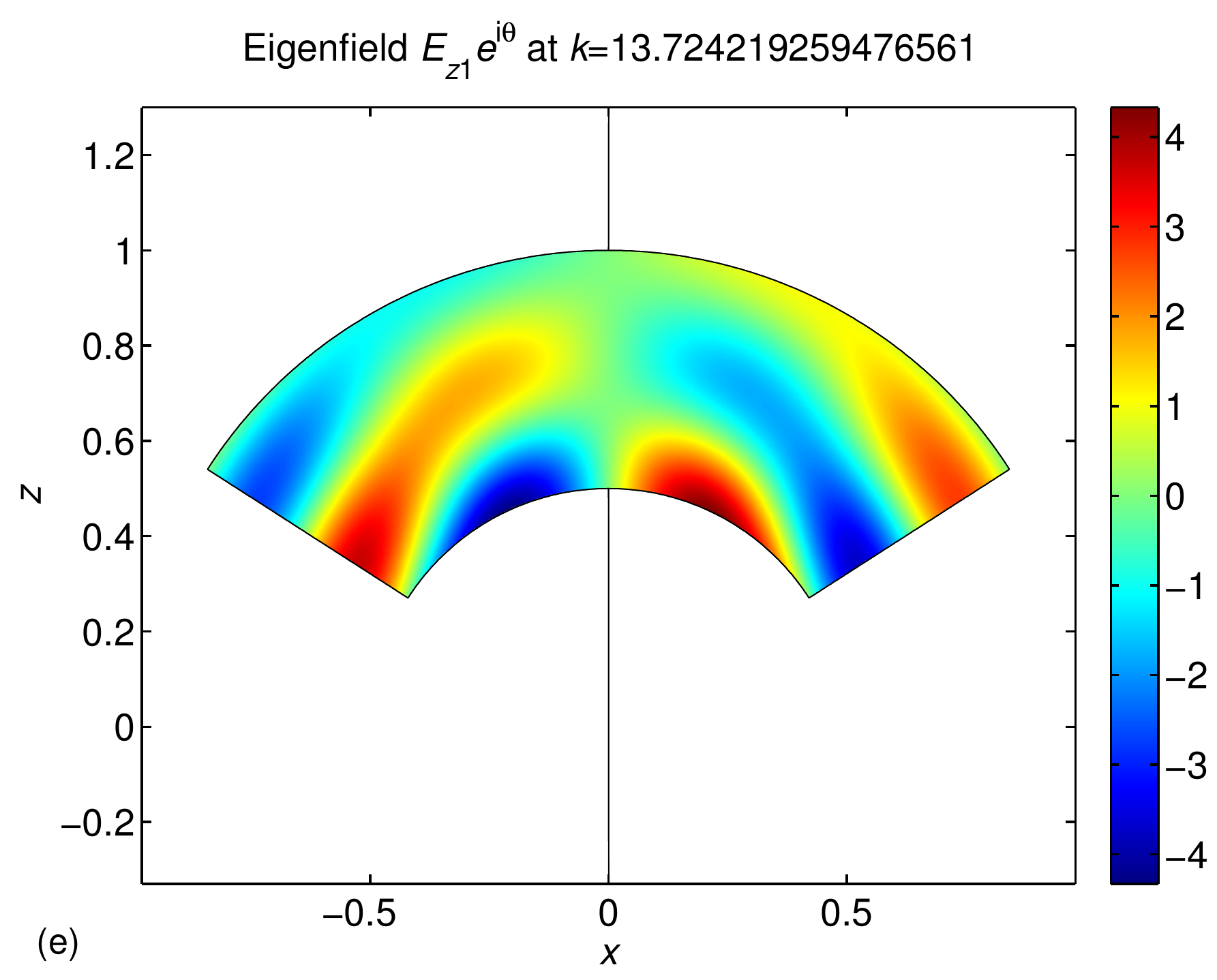}
\includegraphics[height=49mm]{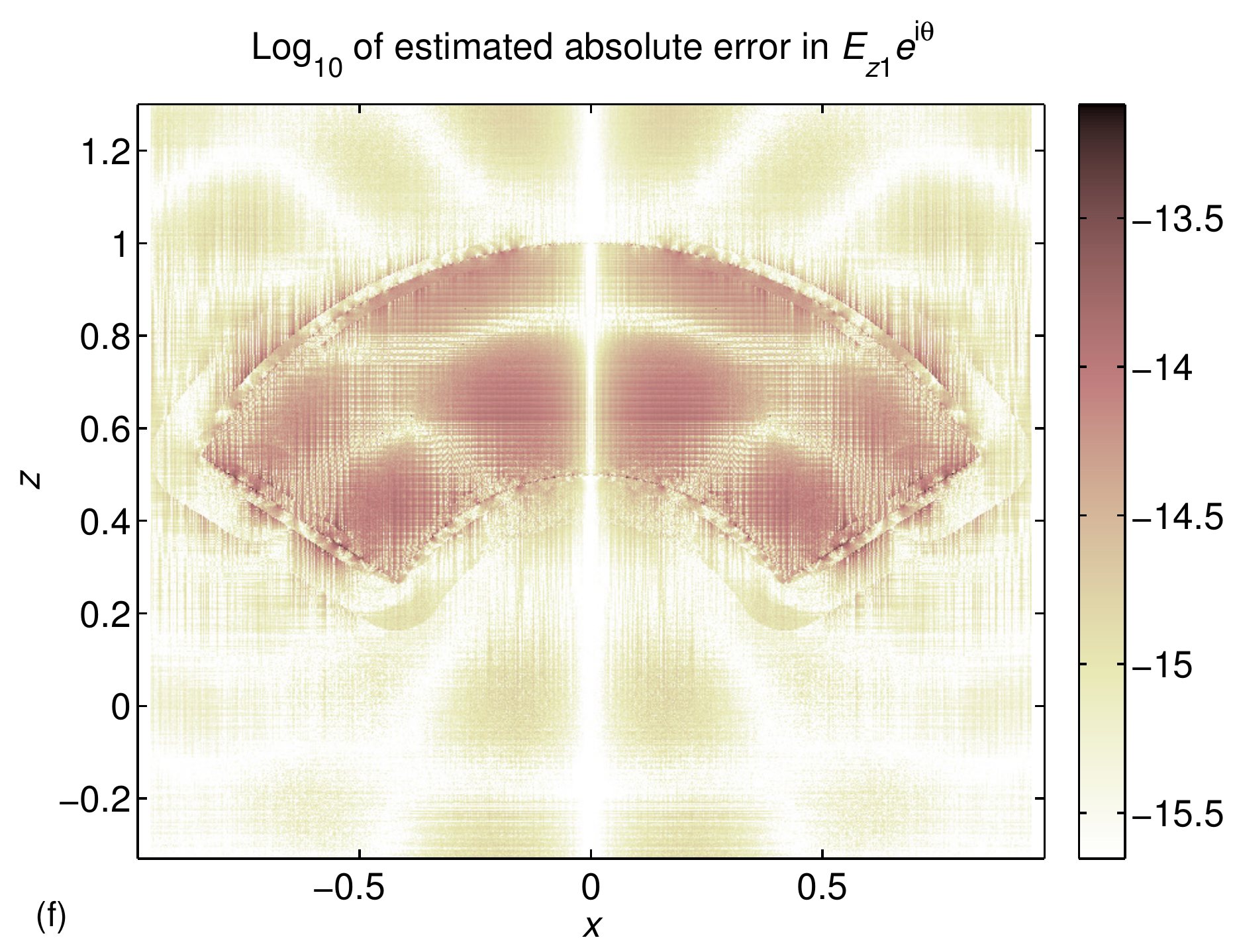}
\caption{\sf The electric eigenfield at $k_{1,10}=13.724219259476561$ 
  for the cavity with $\gamma$ as in~(\ref{eq:param1}). Left: field
  maps in the $xz$-plane ($\theta=0,\pi$) of (a) ${\rm
    Im}\left\{E_{\rho 1}(r)e^{{\rm i}\theta}\right\}$, (c) ${\rm
    Re}\left\{E_{\theta 1}(r)e^{{\rm i}\theta}\right\}$, and (e) ${\rm
    Im}\left\{E_{z 1}(r)e^{{\rm i}\theta}\right\}$. The null-field
  outside the cavity is omitted. Right: (b), (d), and (f) show
  $\log_{10}$ of the pointwise absolute difference between the field
  maps of our scheme and those from the semi-analytic solution. See
  Section~\ref{sec:fysikalisk} for an explanation of how field maps
  relate to physical time-domain fields.}
\label{fig:cone}
\end{figure}

\subsection{Comparison with solution in semi-analytic form}

The codes are first verified for $V$ being the intersection of a cone
with half opening angle of one radian and a spherical shell with outer
radius one and inner radius 0.5. The generating curve $\gamma$ is
parameterized as
\begin{equation}
r(t)=\left\{
\begin{array}{ll}
0.5\left(\sin(2t+2),\cos(2t+2)\right)\,,\quad & t\in[-1,-0.5]\,,\\
(t+1)\left(\sin(1),\cos(1)\right)\,,    \quad & t\in[-0.5,0]\,,\\
\left(\sin(1-t),\cos(1-t)\right)\,,     \quad & t\in[0,1]\,,
\end{array}
\right.
\label{eq:param1}
\end{equation}
and has non-reentrant corners with vertices at
$\gamma_1=0.5\left(\sin(1),\cos(1)\right)$ and
$\gamma_2=\left(\sin(1),\cos(1)\right)$. This cavity is an excellent
test geometry since, while not trivial, it allows for semi-analytic
solutions to the original system~(\ref{eq:PDE1}),~(\ref{eq:PDE2}),
and~(\ref{eq:BC}). The normalized eigenfields $\vec E_n(\vec r)$ are
expressed in regular and irregular spherical vector waves,
see~\cite[Section 2.2]{BosKriStr91}, that are modified to satisfy the
boundary condition (\ref{eq:BC}). The major modifications are that the
associated Legendre functions $P_\nu^n(z/\vert\vec r\vert)$ and
spherical Bessel and Neumann functions, $j_\nu(k\vert\vec r\vert)$ and
$y_\nu(k\vert\vec r\vert)$ in the vector waves, have indices $\nu$ and
wavenumbers $k$ that are solutions to transcendental equations.

Figure~\ref{fig:cone}(a), \ref{fig:cone}(c), and \ref{fig:cone}(e)
show field maps of $E_{\alpha 1}(r)e^{{\rm i}\theta}$,
$\alpha=\rho,\theta,z$, at eigenwavenumber
$k_{1,10}=13.724219259476561$ constructed from Fourier coefficients
$E_{\alpha 1}(r)$ produced by our codes. The eigenwavenumber
corresponds to about 3.7 wavelengths across the generalized diameter
of $V$. The Fourier coefficients are obtained with $n_{\rm pan}=28$
quadrature panels, corresponding to 448 discretization points on
$\gamma$, and they are evaluated at $245000$ points on a Cartesian
grid in the rectangle
\begin{equation}
\Omega_{\rm I}=\left\{r\in\mathbb{R}^2: 0\le\rho\le 0.95\,, 
                            -0.45\le z\le 1.45\right\}\,.
\end{equation}
Only coefficients with $r\in A$ are actually used in
Figure~\ref{fig:cone}(a), \ref{fig:cone}(c), and \ref{fig:cone}(e).
The FFT operations are controlled by the integers $N=141$, $N_{{\rm
    con}1}=24$, and $N_{{\rm con}2}=N_{\rm con}=30$, see
Sections~\ref{sec:discrete} and~\ref{sec:particul}. The densities
$\varrho_{{\rm s}0}(r)$, $J_{\tau 0}(r)$, and $J_{\theta 0}(r)$ are
bounded and the number of panel subdivisions used by the RCIP method
for resolution close to $\gamma_1$ and $\gamma_2$ is chosen as
$n_{{\rm sub}1}=n_{{\rm sub}2}=30$, see Section~\ref{sec:basics}. The
solution time is around 13 seconds and the time required to evaluate
the coefficient vector $\left(E_{\rho 1}(r),E_{\theta 1}(r),E_{z
    1}(r)\right)$ is, on average, 0.002 seconds per point $r$.

Figure~\ref{fig:cone}(b), \ref{fig:cone}(d), and \ref{fig:cone}(f)
show $\log_{10}$ of the absolute difference between the field maps
produced by our codes and the field maps obtained from the
semi-analytic solution. Here all coefficients with $r\in\Omega_{\rm
  I}$ are used. When obtaining the semi-analytic solution, the
eigenwavenumber is evaluated to machine precision and the eigenfields
to almost machine precision using a combination of {\sc Matlab} with
extended precision and {\sf Maple}. The semi-analytic solution at
points $r$ outside $A\cup\gamma$ is taken as $E_{\alpha 1}(r)=0$,
compare~(\ref{eq:comp13F}). One can conclude that, in this example,
our codes give coefficients $E_{\alpha 1}(r)$ that are pointwise
accurate to at least 13 digits and an eigenwavenumber that is accurate
to machine precision.

\subsection{The one cell elliptic cavity}

Our remaining numerical examples pertain to the cavity depicted in
Figure~\ref{fig:geometry} which, in particle accelerator terminology,
is known as a one cell elliptic cavity. The generating curve $\gamma$
is parameterized as
\begin{equation}
r(t)=\left\{
\begin{array}{ll}
\left(\pi+t,-1-\pi/4\right)\,,             \quad & t\in[-\pi,-3\pi/4]\,,\\
\left(\pi/4,-1+\pi/2+t\right)\,,           \quad & t\in[-3\pi/4,-\pi/2]\,,\\
\left(\pi/4+\cos(t),\sin(t)\right)\,,\quad & t\in[-\pi/2,\pi/2]\,,\\
\left(\pi/4,1-\pi/2+t\right)\,,            \quad & t\in[\pi/2,3\pi/4]\,,\\
\left(\pi-t,1+\pi/4\right)\,,              \quad & t\in[3\pi/4,\pi]\,,
\end{array}
\right.
\label{eq:param2}
\end{equation}
and has corner vertices at $\gamma_1=(\pi/4,-1-\pi/4)$,
$\gamma_2=(\pi/4,-1)$, $\gamma_3=(\pi/4,1)$, and
$\gamma_4=(\pi/4,1+\pi/4)$. The corners at $\gamma_2$ and $\gamma_3$
are reentrant. The number of panel subdivisions in the RCIP method is
chosen as $n_{{\rm sub}1}=n_{{\rm sub}4}=30$ and $n_{{\rm
    sub}2}=n_{{\rm sub}3}=140$ in all examples. The integer $N$,
controlling the FFT operations when $r$ and $r'$ are far from each
other, is by default chosen as $N=4n_{\rm pan}+n$. The {\it estimated
  pointwise absolute error} in a given computed field map is based on
a comparison with a more resolved map obtained with 50 per cent more
quadrature panels on $\gamma$. Fourier coefficients are evaluated on
Cartesian grids in rectangles, most often chosen as
\begin{equation}
\Omega_{\rm II}=\left\{r\in\mathbb{R}^2: 0\le\rho\le 2\,, 
                               -2\le z\le 2\right\}\,.
\label{eq:ellkaO}
\end{equation}

Superconducting elliptic cavities are common in linear accelerators
for protons. They are to be used in a projected Superconducting Proton
Linear accelerator (SPL) at CERN and in the linear accelerator for the
European Spallation Source (ESS) that is currently under construction
in Lund, Sweden. In the design of elliptic cavities it is important to
determine several quantities to high accuracy. For the fundamental
eigenfield that accelerates the protons, one needs to evaluate the
maximum normalized electric and magnetic fields on the surface and the
normalized electric field on the symmetry axis. The eigenfields with
azimuthal indices $n=0$ and $n=1$ have non-zero field components on
the symmetry axis which cause them to interact with the beam of
particles. A large number of these eigenfields have to be determined
in order to assess their effect on the beam. The three numerical
examples we present for the elliptic cavity have $n=0$ and $n=1$ and
eigenwavenumbers that are relevant for particle accelerators.

\begin{figure}[!t]
  \centering 
  \includegraphics[height=55mm]{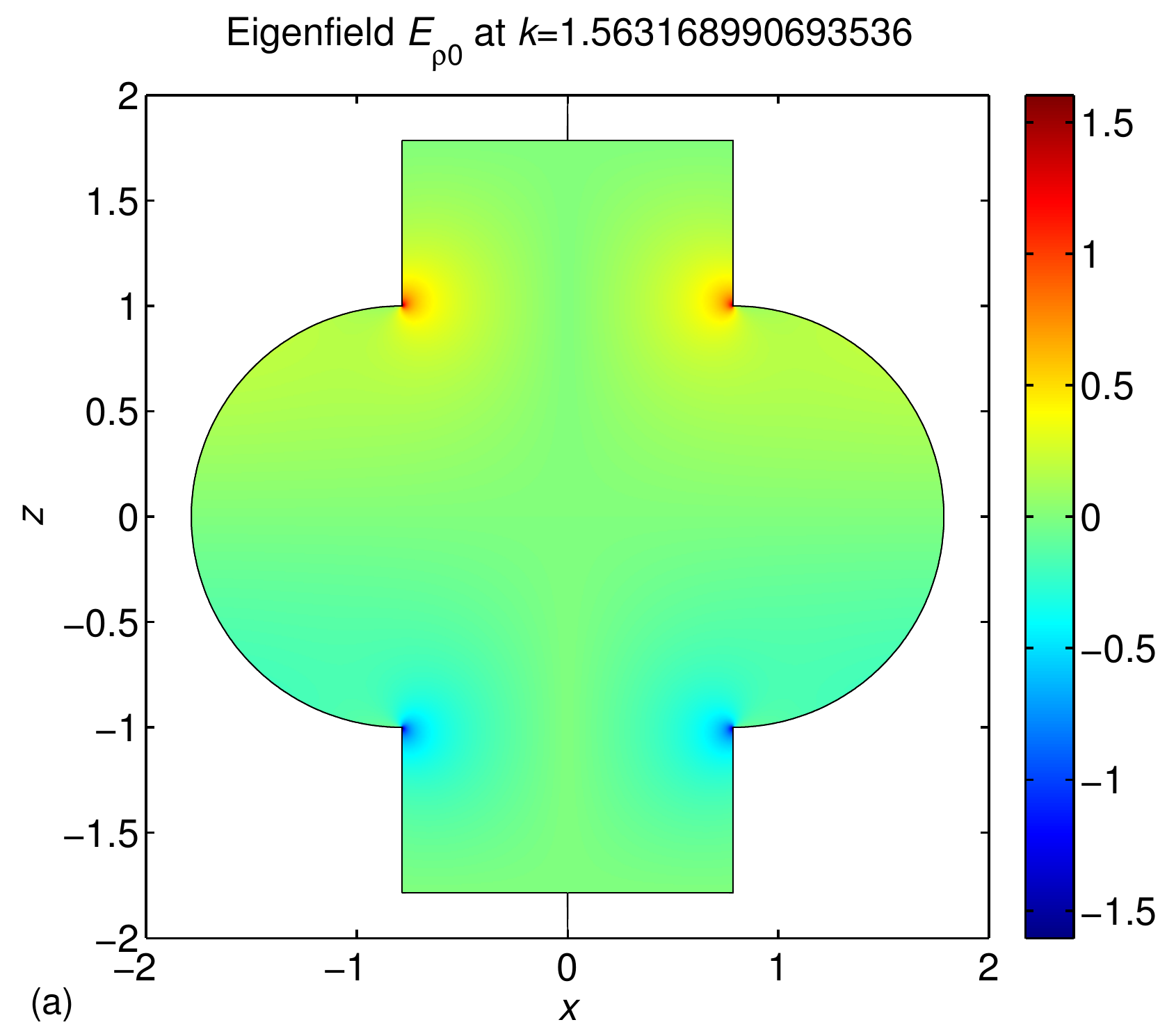}
  \includegraphics[height=55mm]{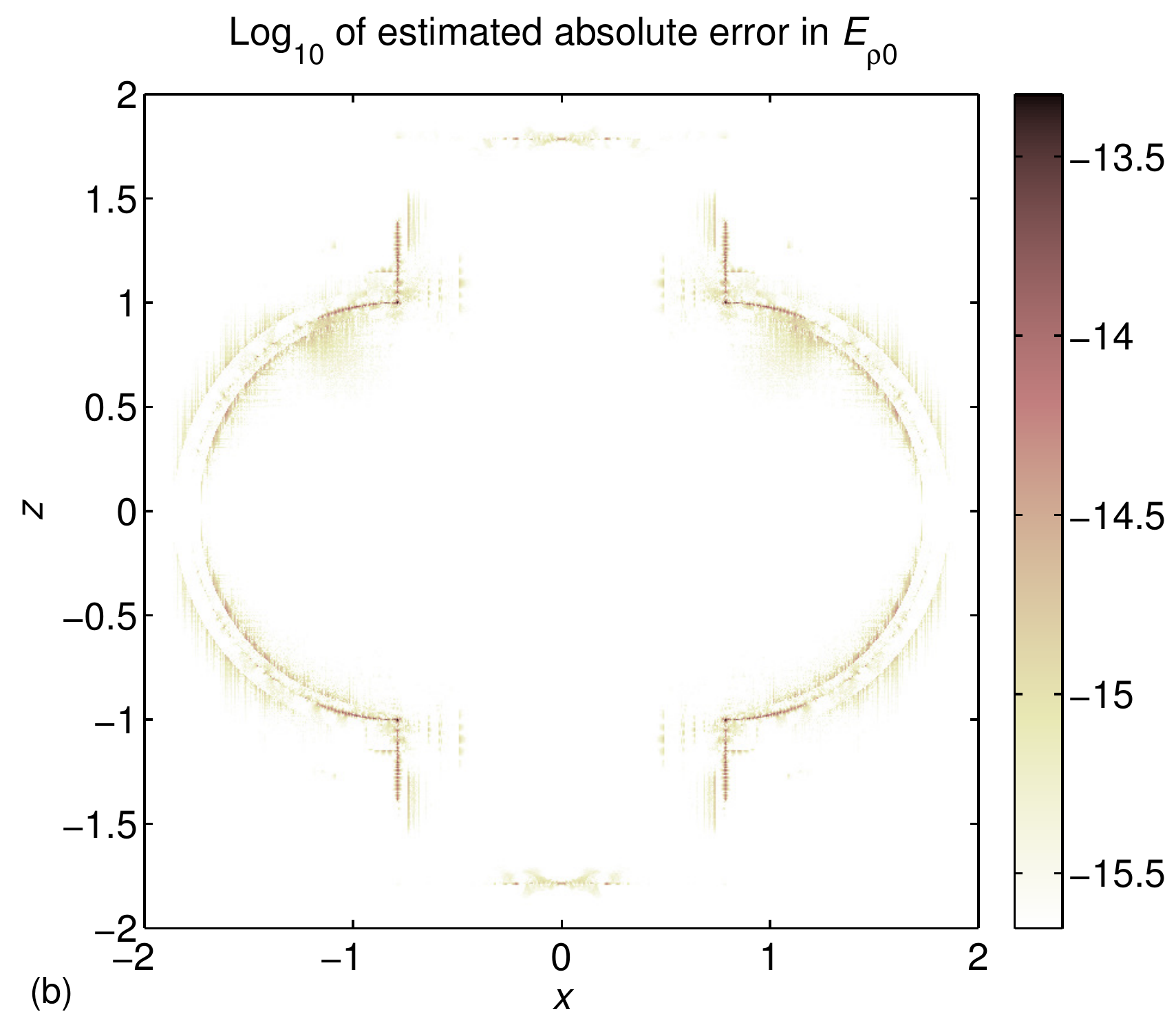}
  \includegraphics[height=55mm]{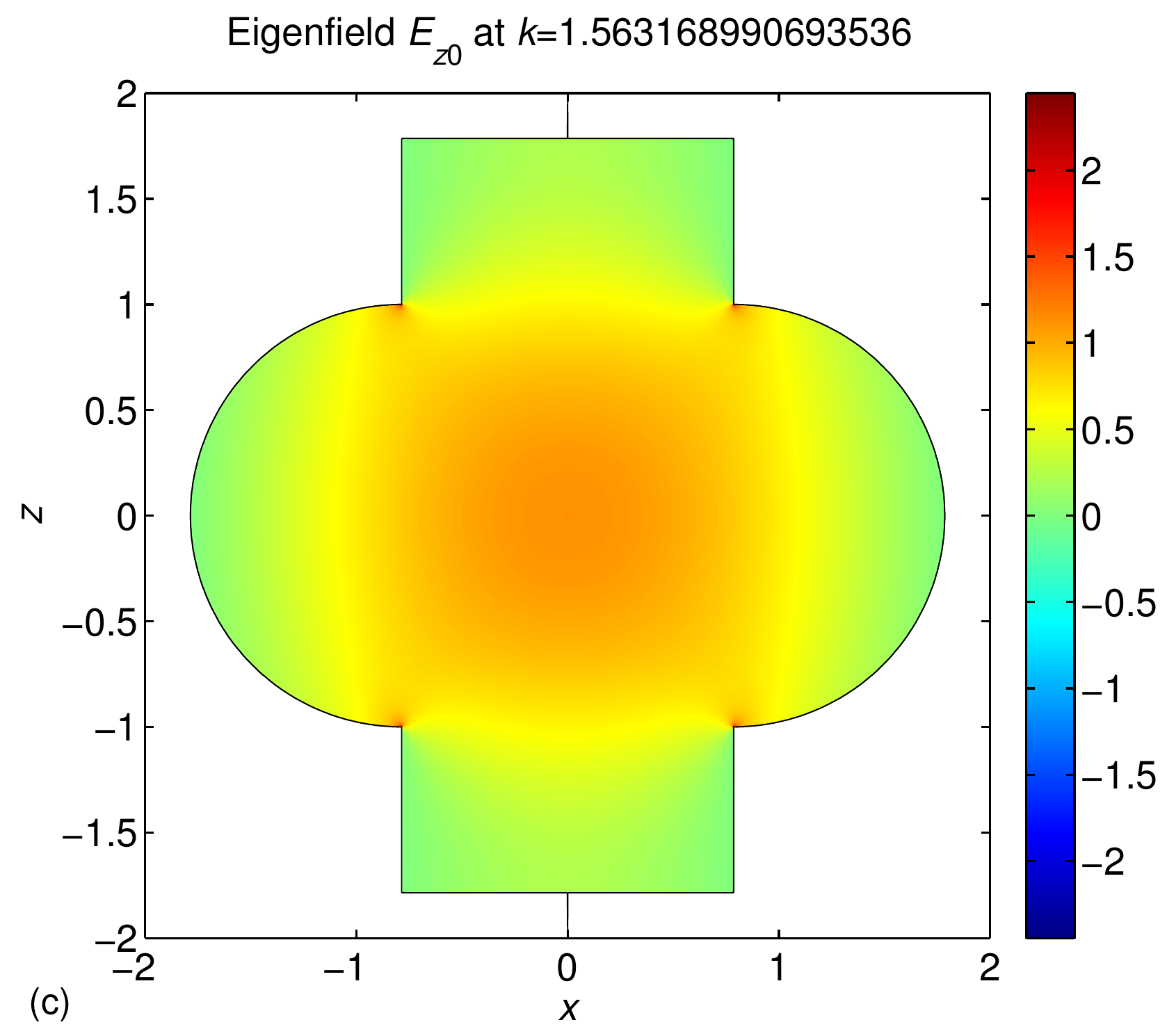}
  \includegraphics[height=55mm]{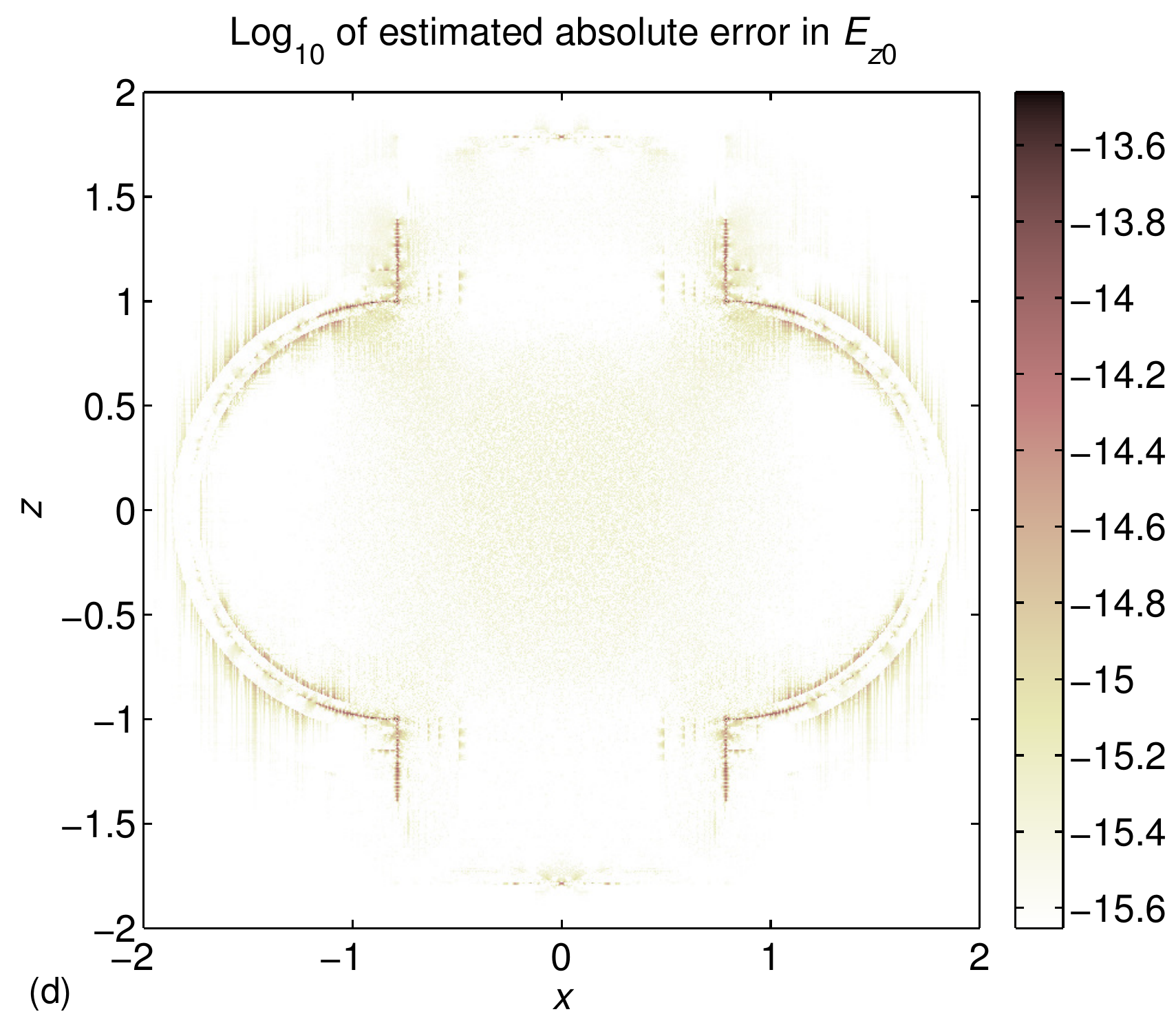}
\caption{\sf The fundamental electric eigenfield for the elliptic
  cavity with $\gamma$ as in~(\ref{eq:param2}). Left: field maps in
  the $xz$-plane of (a) ${\rm Im}\left\{E_{\rho 0}(r)\right\}$ and (c)
  ${\rm Im}\left\{E_{z 0}(r)\right\}$. Right: $\log_{10}$ of estimated
  pointwise absolute error.}
\label{fig:elk1}
\end{figure}

\subsection{The fundamental mode}

The fundamental electric eigenfield is the eigenfield with the lowest
eigenwavenumber. For the elliptic cavity with $\gamma$ as
in~(\ref{eq:param2}) it has eigenwavenumber
$k_{0,1}=1.5631689906935362$, corresponding to 0.97 wavelengths across
the generalized diameter of $V$. Figure~\ref{fig:elk1}(a) and
\ref{fig:elk1}(c) show field maps of $E_{\rho 0}(r)$ and $E_{z 0}(r)$
as computed with our scheme. The map of $E_{\theta 0}(r)$ is zero and
therefore omitted. The Fourier coefficients are obtained with $n_{\rm
  pan}=32$ quadrature panels, corresponding to 512 discretization
points on $\gamma$, and they are evaluated at 245000 points on a grid
in $\Omega_{\rm II}$. The FFT operations, for $r$ and $r'$ close, are
controlled by $N_{{\rm con}1}=N_{{\rm con}4}=12$, $N_{{\rm
    con}2}=N_{{\rm con}3}=14$ and $N_{\rm con}=16$. The solution time
is around 16 seconds and the time required to evaluate the coefficient
vector $\left(E_{\rho 0}(r),E_{z 0}(r)\right)$ is, on average, 0.003
seconds per point $r$.

Figure~\ref{fig:elk1}(b) and \ref{fig:elk1}(d) show $\log_{10}$ of the
estimated pointwise absolute error in Figure~\ref{fig:elk1}(a) and
\ref{fig:elk1}(c). At this low eigenwavenumber the estimated accuracy
is quite exceptional. The solver delivers 15 accurate digits except at
points close to $\gamma$.

Note that $E_{z 0}(r)$ is strong along the symmetry axis. This
explains why the fundamental mode is used for acceleration of charged
particles. At the vertices of the reentrant corners both $E_{\rho
  0}(r)$ and $E_{z 0}(r)$ diverge as $\xi^{-1/3}$, see
Section~\ref{sec:asymp}. In the design of de facto elliptic cavities
in accelerators, sharp reentrant cell- and iris edges are avoided. On
the other hand, there are sharp reentrant edges where the beam pipe is
attached to the cavity and it is therefore important that a solver can
handle all sorts of sharp edges.  We have omitted the beam pipe in
order to keep the model simple.

\begin{figure}[!t]
  \centering 
\noindent\makebox[\textwidth]{
\begin{minipage}{1.13\textwidth}
\hspace{1mm}
\includegraphics[height=60mm]{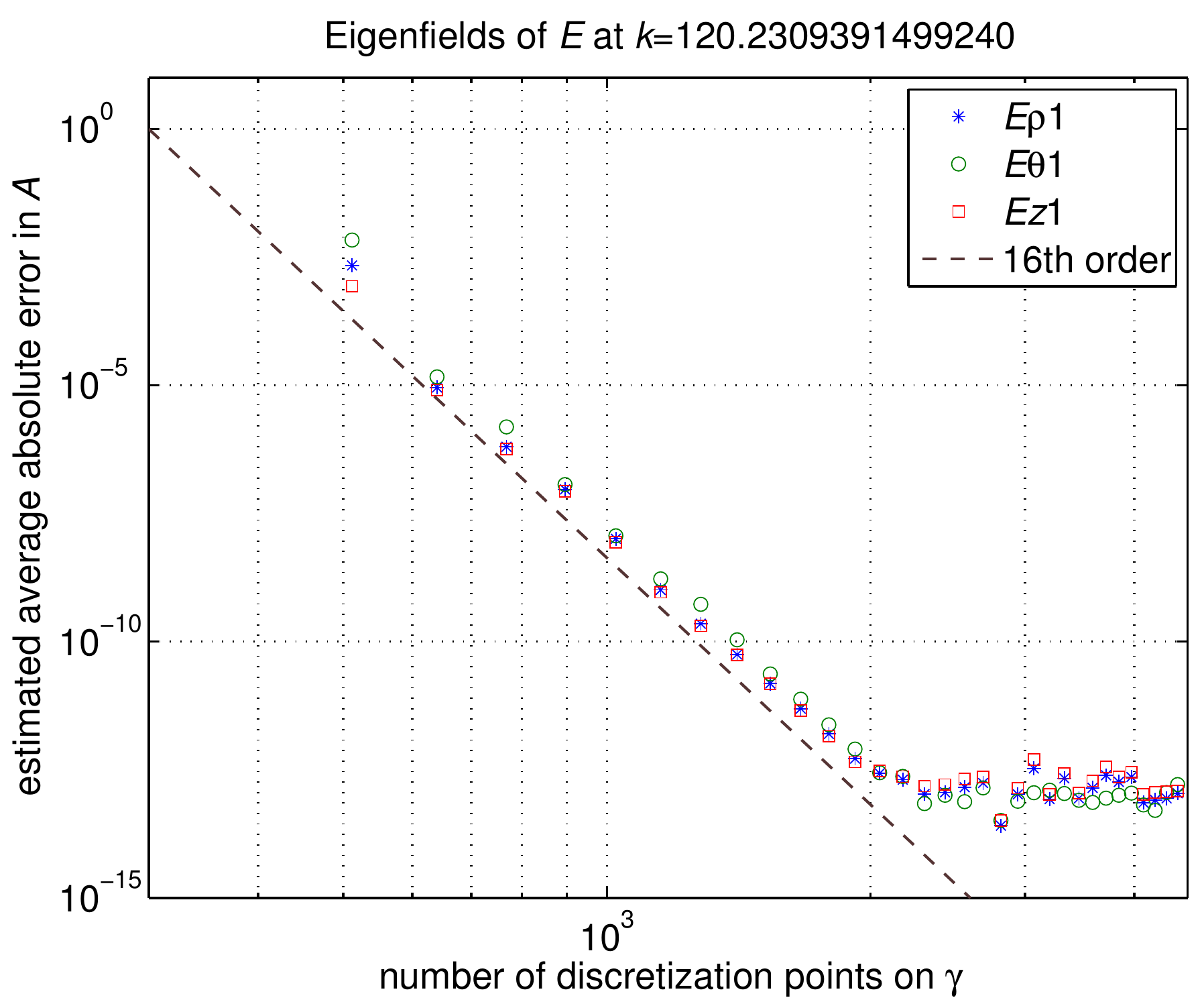}
\includegraphics[height=60mm]{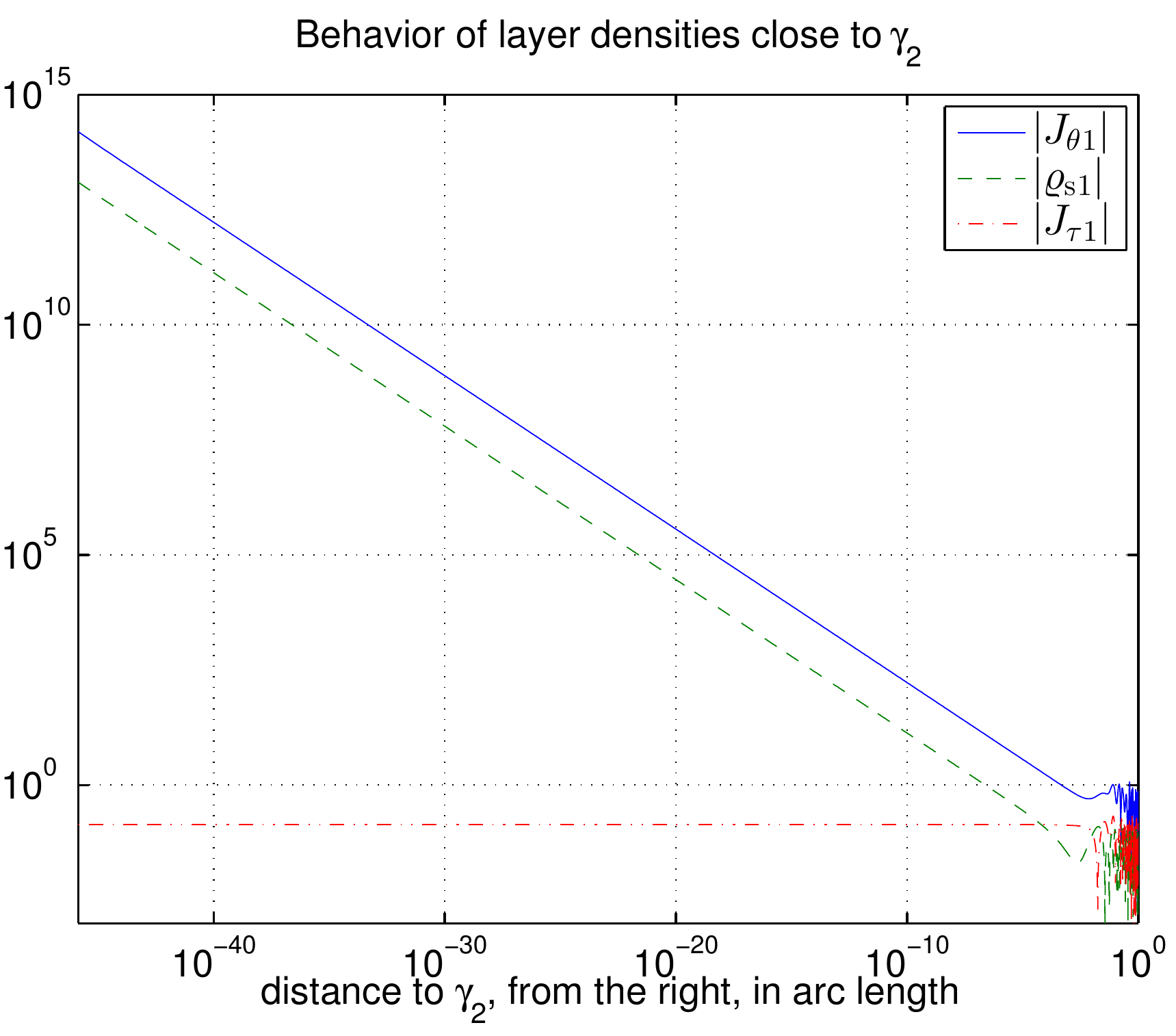}
\end{minipage}}
\caption{\sf Left: convergence of $E_{\rho 1}(r)$, $E_{\theta 1}(r)$, 
  and $E_{z 1}(r)$ with $\gamma$ given by~(\ref{eq:param2}) and at
  eigenwavenumber $k_{1,9928}=120.2309391499240$. The estimated
  average pointwise absolute error in $A$ has converged to less than
  $10^{-12}$ at $1920$ discretization points on $\gamma$,
  corresponding to 16 points per wavelength along $\gamma$.  Right:
  behavior of $\varrho_{{\rm s}1}(r)$, $J_{\tau 1}(r)$, and $J_{\theta
    1}(r)$ close to $\gamma_2$.}
\label{fig:conv120}
\end{figure}

\begin{figure}[!t]
\centering
\includegraphics[height=55mm]{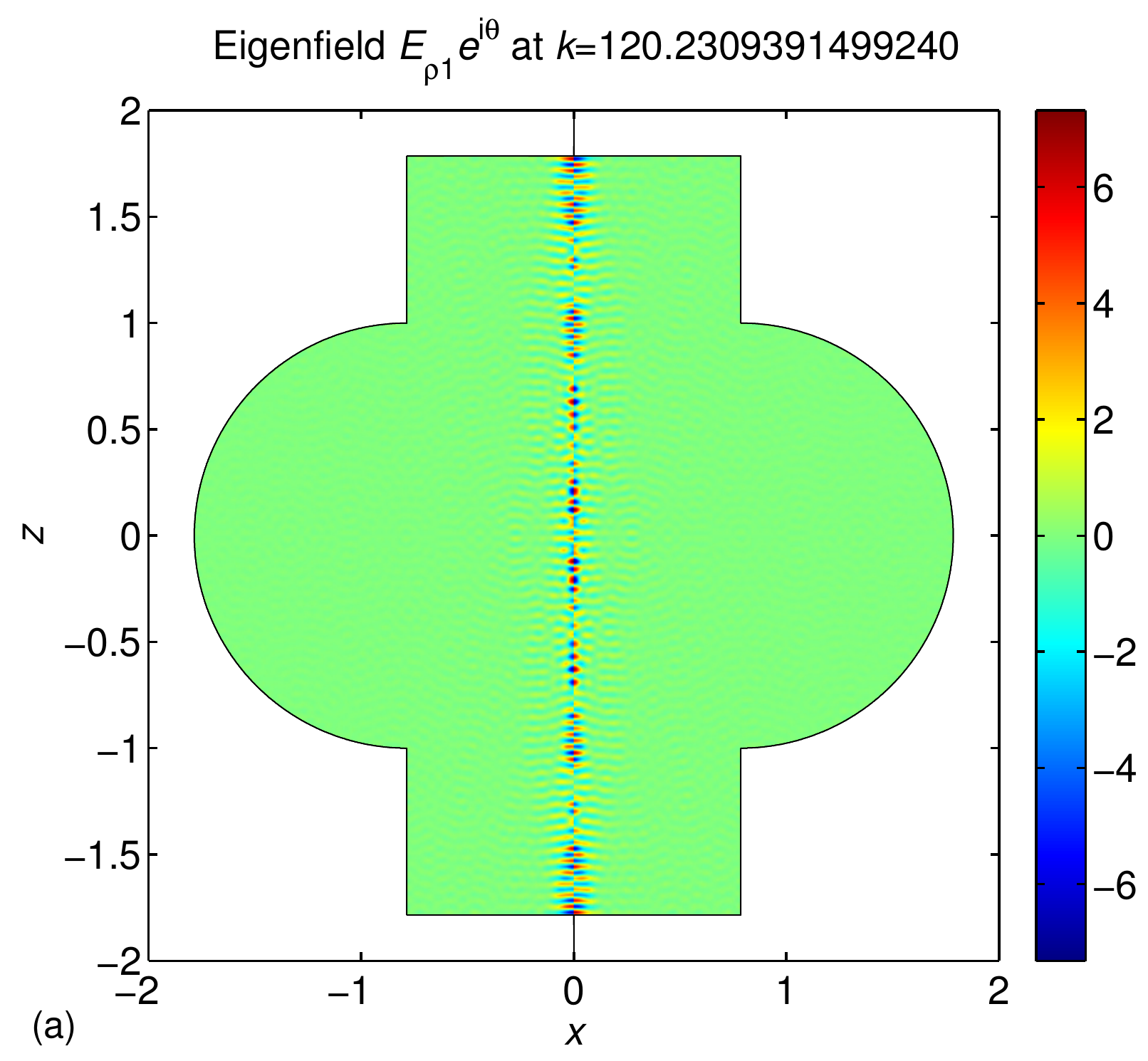}
\includegraphics[height=55mm]{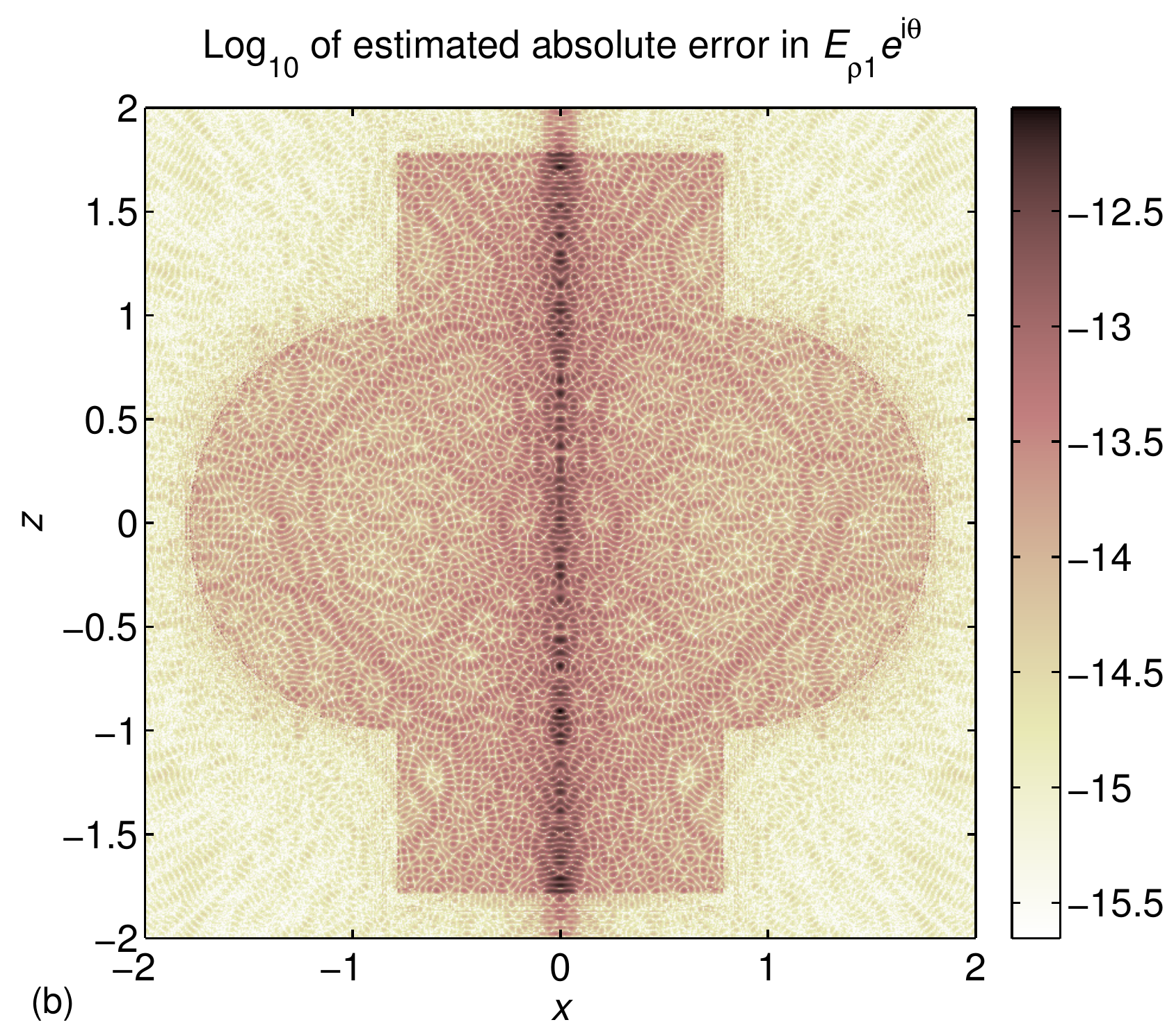}
\includegraphics[height=55mm]{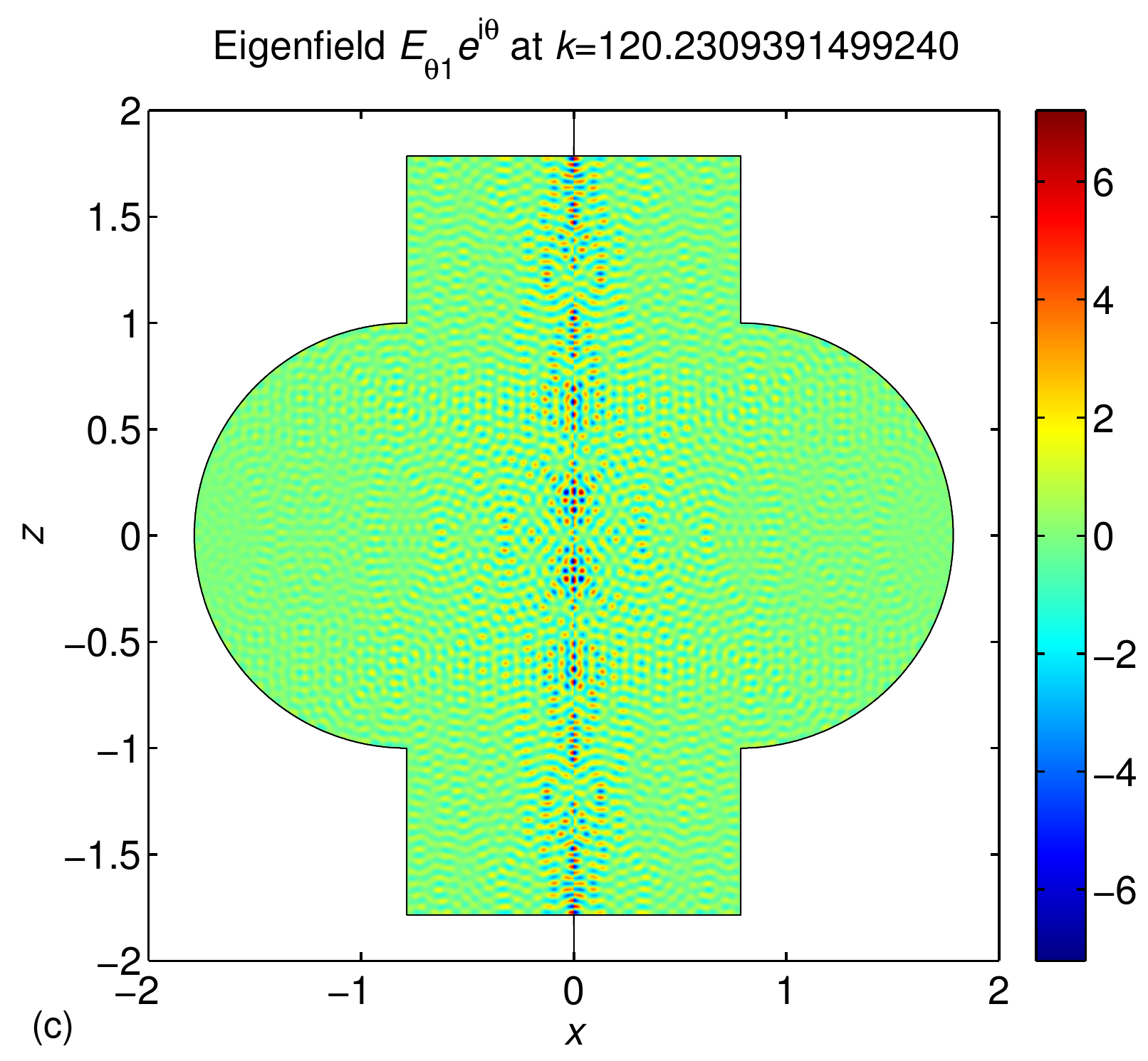}
\includegraphics[height=55mm]{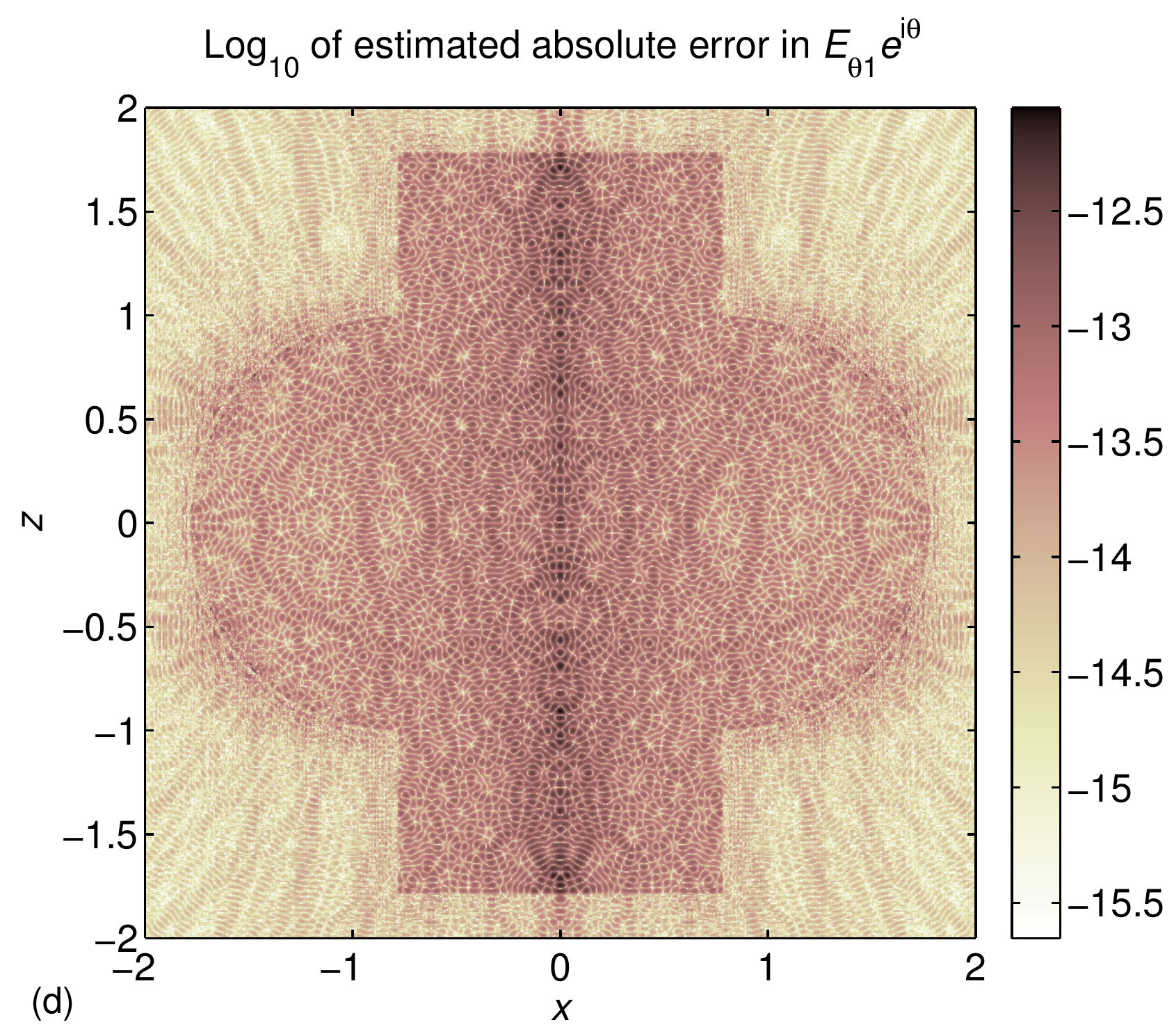}
\includegraphics[height=55mm]{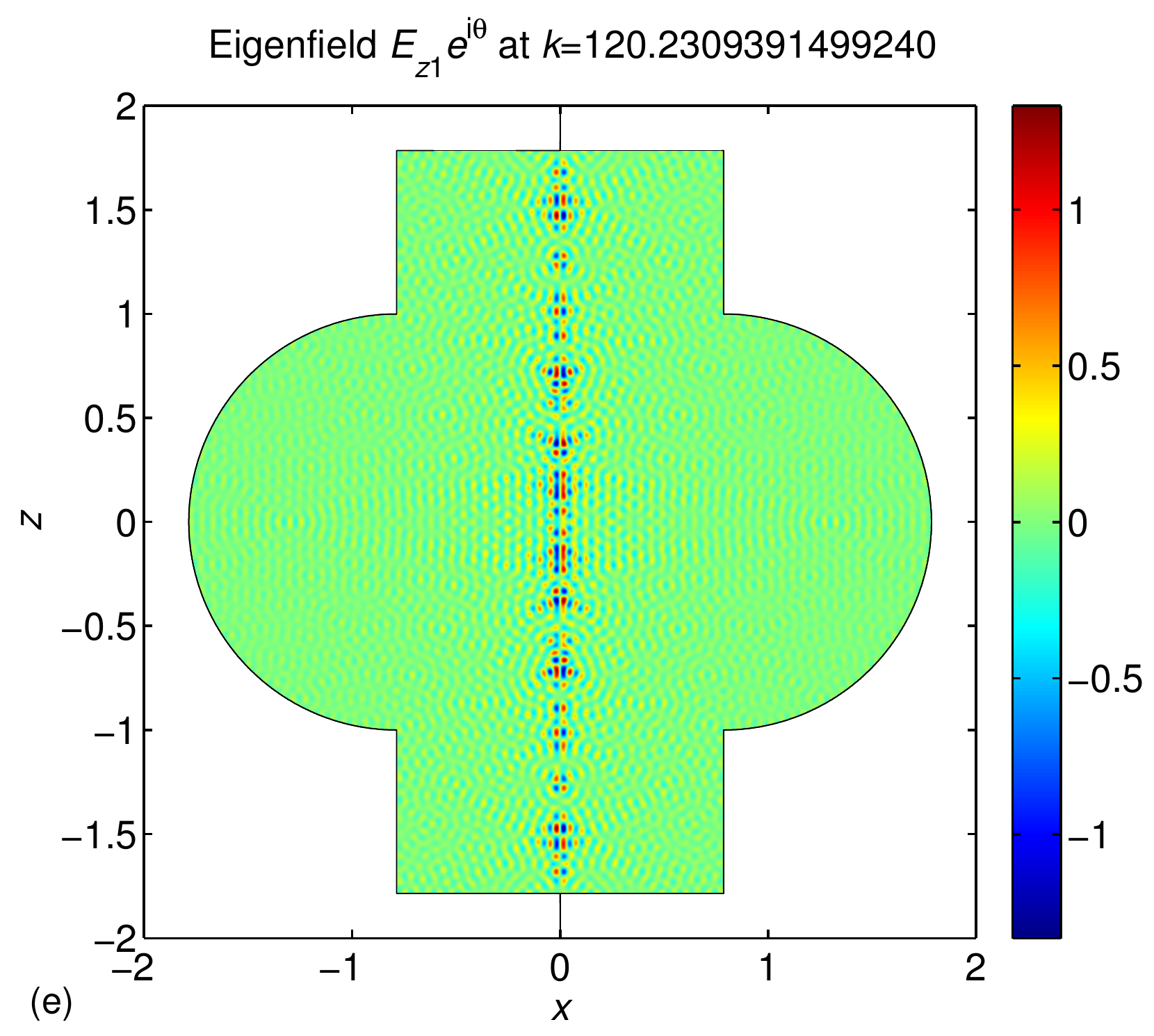}
\includegraphics[height=55mm]{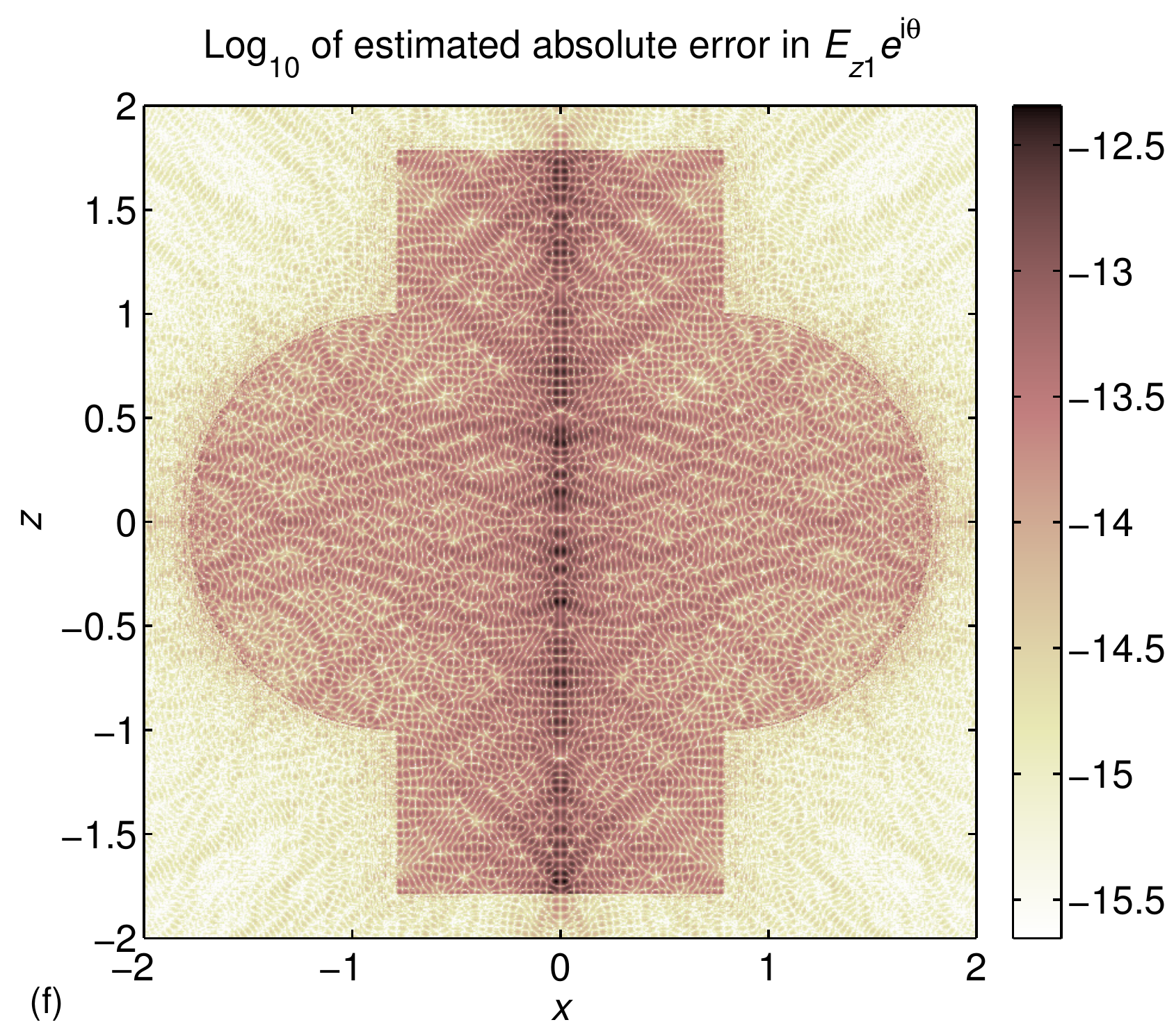}
\caption{\sf The electric eigenfield for the elliptic cavity with
  $\gamma$ as in~(\ref{eq:param2}) and with
  $k_{1,9928}=120.2309391499240$. Left: field maps in the $xz$-plane
  of (a) ${\rm Im}\left\{E_{\rho 1}(r)e^{{\rm i}\theta}\right\}$, (c)
  ${\rm Re}\left\{E_{\theta 1}(r)e^{{\rm i}\theta}\right\}$, and (e)
  ${\rm Im}\left\{E_{z 1}(r)e^{{\rm i}\theta}\right\}$. Right:
  $\log_{10}$ of estimated pointwise absolute error.}
\label{fig:elk120}
\end{figure}

\subsection{A convergence study}

The next example is the eigenfield with
$k_{1,9928}=120.2309391499240$. Despite a generalized diameter of the
elliptic cavity that now corresponds to around 75 wavelengths, our
solver maintains 16th order convergence and high achievable accuracy,
as seen in the left image of Figure \ref{fig:conv120}. The FFT
operations are in this study controlled by $N=\max\{300,4n_{\rm
  pan}+n\}$, $N_{{\rm con}1}=N_{{\rm con}4}=127$, $N_{{\rm
    con}2}=N_{{\rm con}3}=137$, and $N_{\rm con}=259$. The Fourier
coefficients are evaluated at 45000 points on a grid in $\Omega_{\rm
  II}$ and have converged to more than 12 digits already at 16 points
per wavelength along $\gamma$, which is marginally better than in a
similar study for the eigenfield with $k_{1,2460}=60.21392380136615$
(not shown). We conclude that there are no signs of any pollution
effect~\cite{Babu97} at these wavenumbers.

The right image of Figure \ref{fig:conv120} reveals that at the
reentrant corners both $\varrho_{{\rm s}1}(r)$ and $J_{\theta 1}(r)$
diverge as $\xi_{\rm t}^{-1/3}$, whereas $J_{\tau 1}(r)$ is bounded.
These asymptotics are in accordance with~(\ref{eq:asymp}).

Figure~\ref{fig:elk120} shows fully converged field maps of $E_{\alpha
  1}(r)e^{{\rm i}\theta}$, $\alpha=\rho,\theta,z$, obtained with 2816
discretization points on $\gamma$, along with $\log_{10}$ of estimated
pointwise absolute error. The solution time is around 240 seconds and
the time required to evaluate the coefficient vector $\left(E_{\rho
    1}(r),E_{\theta 1}(r),E_{z 1}(r)\right)$ is, on average, 0.04
seconds per point $r$. One can see, in the left images, that the
eigenfield is concentrated to a region close to the symmetry axis.
This is typical for eigenfields with small $n$ and large $k$.

It follows from~(\ref{eq:Varrho}) that the normal component of the
coefficient vector has the same (singular) behavior as $\varrho_{{\rm
    s}1}(r)$ along $\gamma$. The amplitude of a singularity in
$\varrho_{{\rm s}1}(r)$ in a reentrant corner is often small at large
eigenwavenumbers. As seen in the right image of
Figure~\ref{fig:conv120}, it may become visible first at a distance
from a corner vertex that is less than one thousandth of the total
arclength. Although the images of Figure~\ref{fig:elk120} use 245000
evaluation points on the grid in $\Omega_{\rm II}$, this is not enough
to detect the singularities in the field maps. This underscores the
importance of being able to zoom regions where singularities might
appear in order to determine their amplitudes.

\begin{figure}[!t]
\centering
\includegraphics[height=55mm]{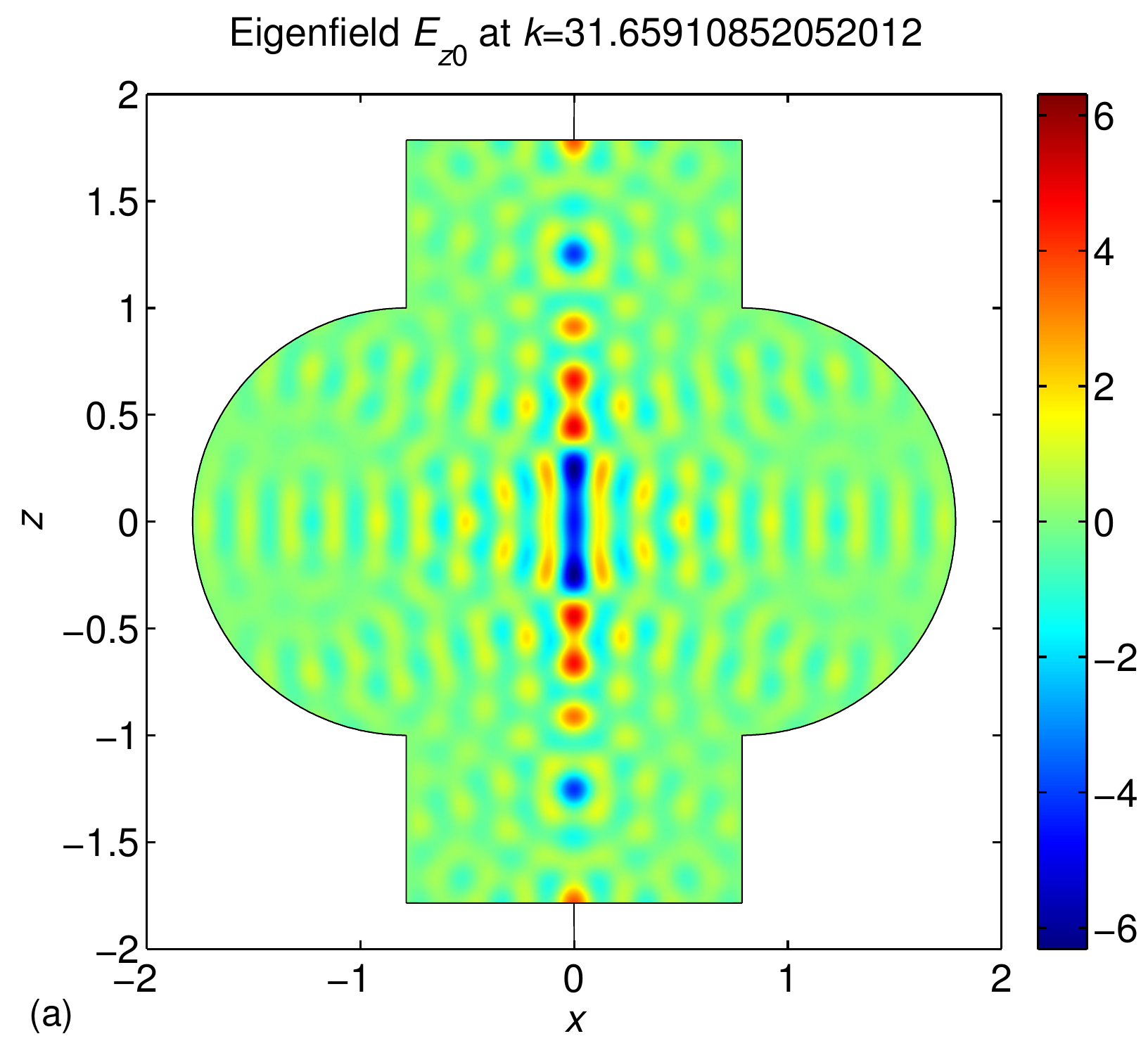}
\includegraphics[height=55mm]{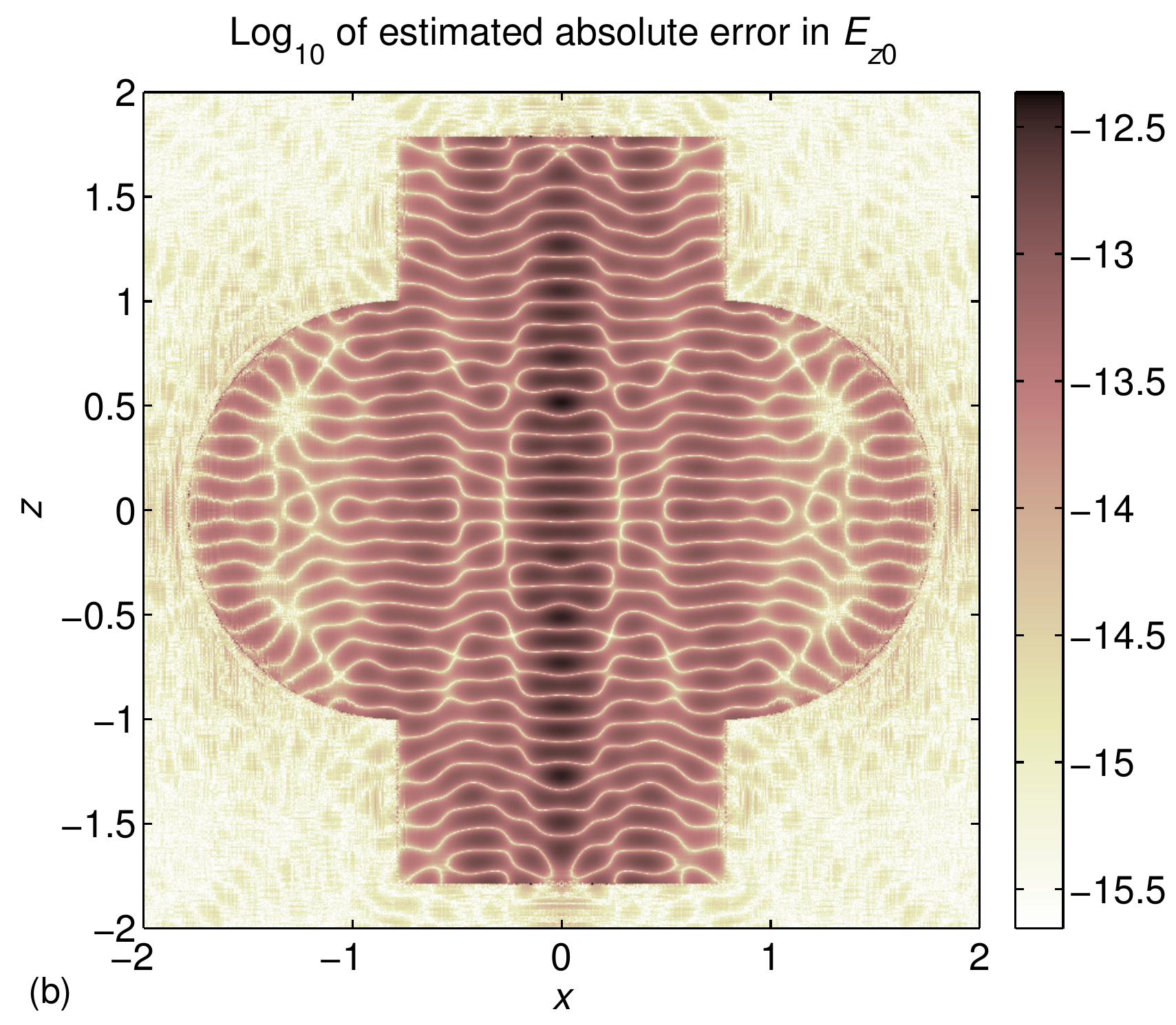}
\includegraphics[height=55mm]{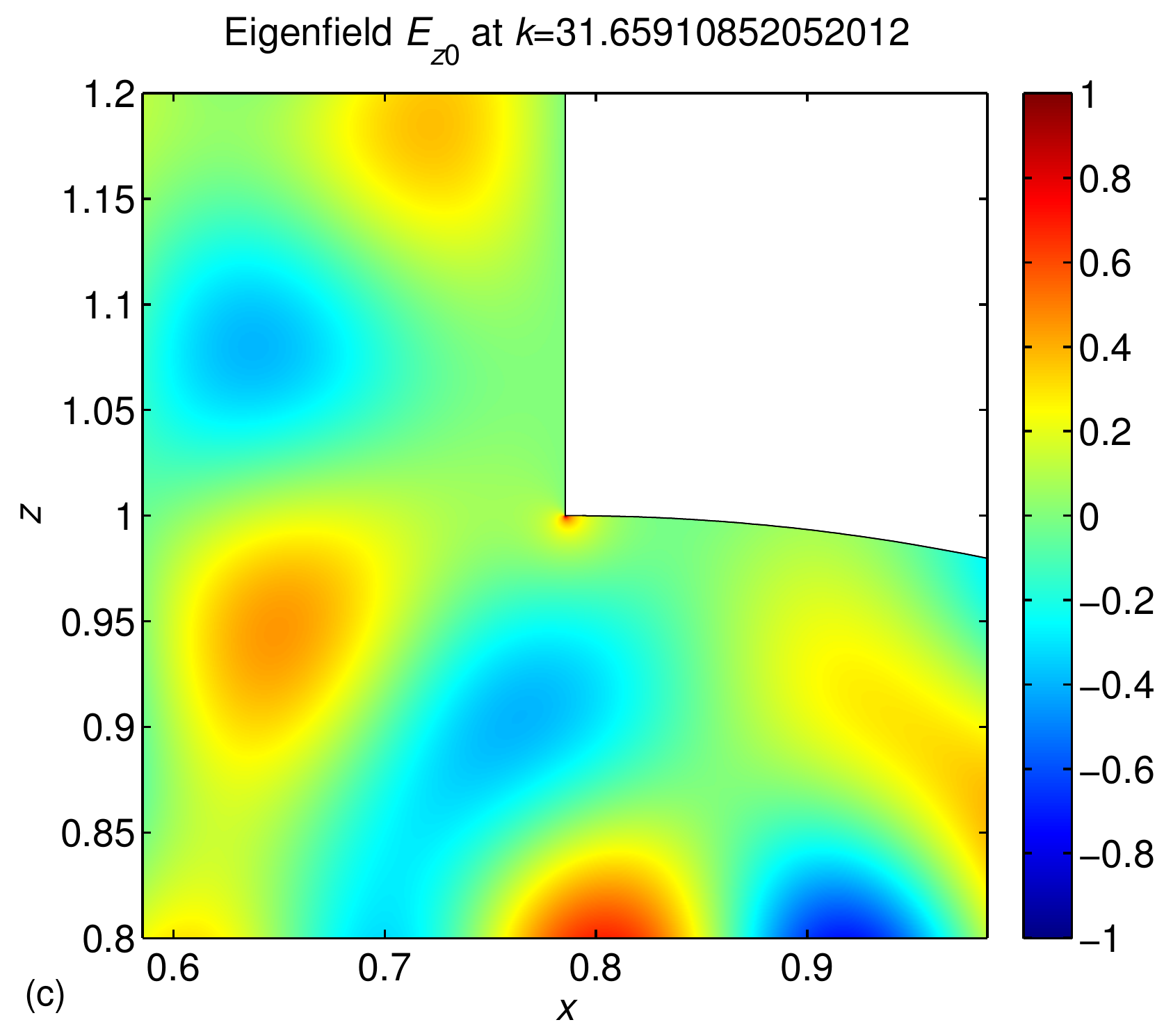}
\includegraphics[height=55mm]{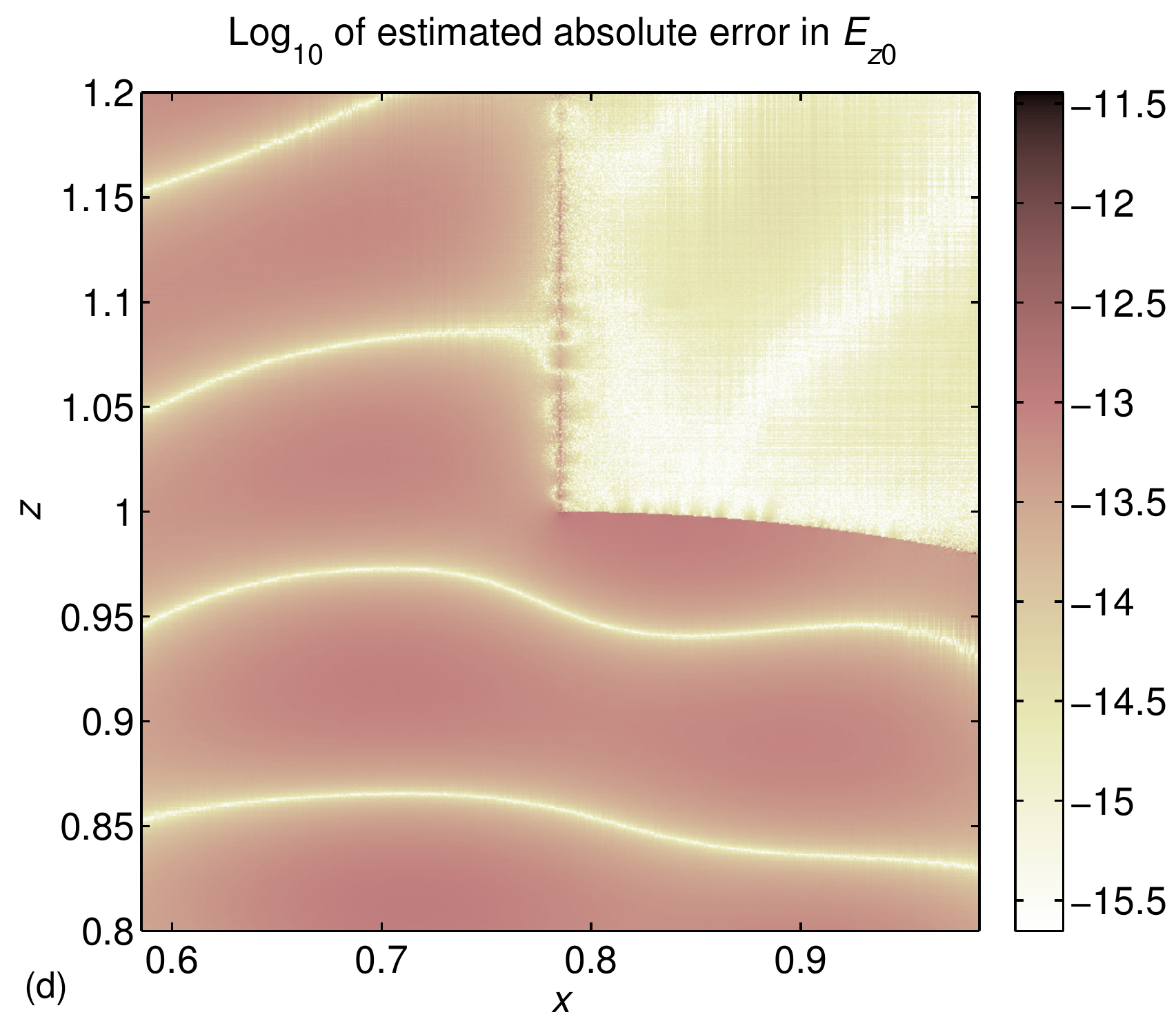}
\includegraphics[height=54mm]{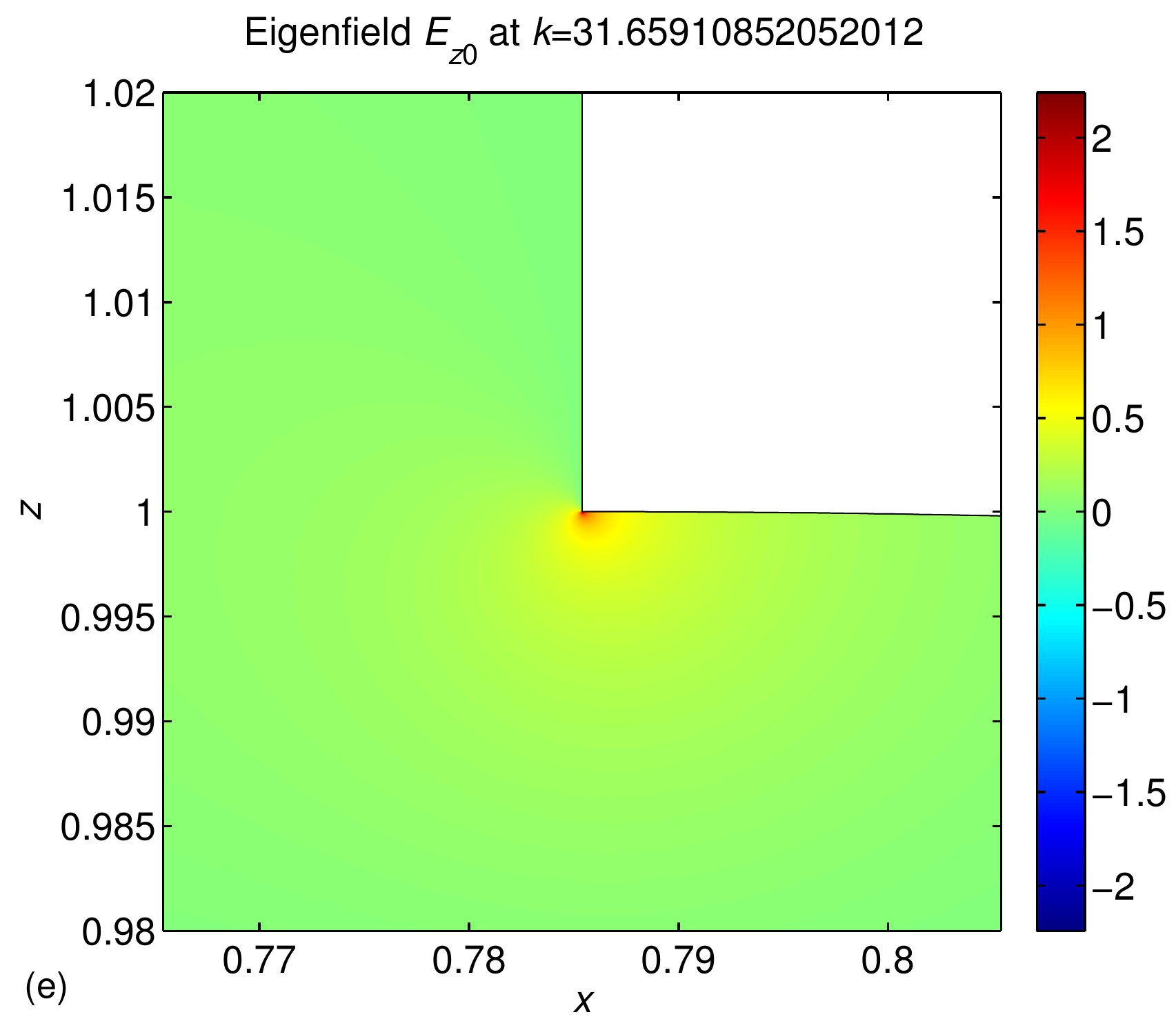}
\includegraphics[height=54mm]{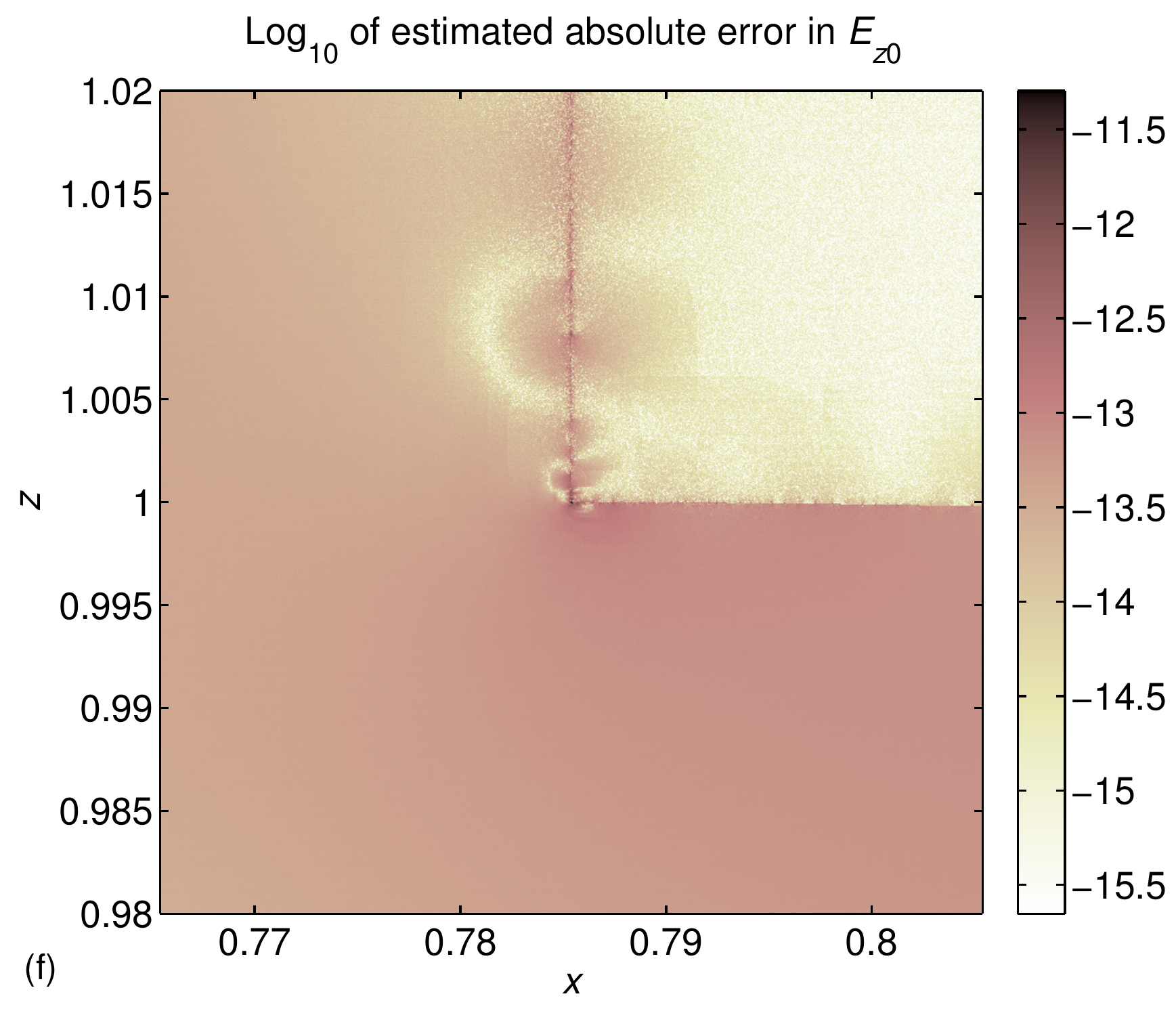}
\caption{\sf The electric eigenfield for the elliptic cavity with 
  $\gamma$ as in~(\ref{eq:param2}) and with
  $k_{0,662}=31.65910852052012$. Left: (a) the field map of ${\rm
    Im}\left\{E_{z 0}(r)\right\}$ in the $xz$-plane, (c) is a ten
  times and (e) a 100 times magnification in the vicinity of the
  corner vertex $\gamma_3$. Right: $\log_{10}$ of estimated pointwise
  absolute error.}
\label{fig:elk31}
\end{figure}

\subsection{Corner zoom}

Our last example is the eigenfield with $k_{0,662}=31.65910852052012$.
This eigenfield is even according to the classification of
Section~\ref{sec:phystime} and has $E_{\theta 0}(r)=0$ in agreement
with Section~\ref{sec:asymp}. Figure~\ref{fig:elk31}(a) shows a field
map of $E_{z 0}(r)$ obtained with 768 discretization points on
$\gamma$, $N_{{\rm con}1}=N_{{\rm con}4}=48$, $N_{{\rm con}2}=N_{{\rm
    con}3}=57$, and $N_{\rm con}=85$ and evaluated at 245000 points on
a grid in $\Omega_{\rm II}$. The solution time is around 37 seconds.

Figure~\ref{fig:elk31}(c) and \ref{fig:elk31}(e) explore the field map
of $E_{z 0}(r)$ when the region around the corner vertex $\gamma_3$ is
magnified first ten times and then 100 times. There are still 768
discretization points on $\gamma$, but field evaluations now take
place at 490000 points on grids in the squares
\begin{equation}
\Omega_\delta=\left\{r\in\mathbb{R}^2: 
\frac{\pi}{4}-\delta\le\rho\le \frac{\pi}{4}+\delta\,, 
1-\delta\le\ z\le 1+\delta\right\}\,,
\label{eq:ellkaz}
\end{equation}
where $\delta=0.2$ or $\delta=0.02$. At $\gamma_3$ (and at $\gamma_2$)
the field map of $E_{z 0}(r)$ exhibits the singularity given
by~(\ref{eq:asymp}). The right images of Figure~\ref{fig:elk31} show
$\log_{10}$ of the estimated pointwise absolute error.

This example emphasizes that high accuracy is vital for the detection
of singular fields. A less accurate solver might neglect strong local
electric fields that can lead to serious electric discharges in a
cavity.

\section{Conclusion and outlook}
\label{sec:conclus}

We have presented a competitive solver for the determination of
normalized electric eigenfields in axially symmetric microwave
cavities with piecewise smooth PEC surfaces. The solver is based on
the following key elements: the interior magnetic field integral
equation, a charge integral equation, a surface integral for
normalization, a high-order convergent Fourier--Nyström discretization
scheme, on-the-fly computation of singular and nearly singular
quadrature rules, and access to high-order surface information.

While our solver determines eigenfields with extraordinary accuracy
for a large range of eigenwavenumbers, we observe that the development
of efficient high-order Nyström schemes for time-harmonic boundary
value problems in piecewise smooth three-dimensional domains is an
active research field. See~\cite{Bremetal15} for a recent example of a
fast solver for the integral equations which model low-frequency
acoustic scattering from curved surfaces.

The computational needs in accelerator technology are extensive and
our solver must be equipped with additional features to become a truly
versatile tool. Our next step is to allow for sources, modeling a
pulsed beam of particles, inside cavities. This makes it possible to
evaluate wakefields generated by beams in accelerators. We foresee
that our solver can be used for benchmarking. It will be able to
evaluate high-frequency parts of wakefield spectra that other solvers
cannot reach and by that it becomes an important complement to
state-of-the-art software such as the CST Particle Studio Wakefield
Solver. We also anticipate other improvements. The present {\sc
  Matlab} implementation places emphasis on high achievable accuracy
and on rapid convergence with respect to the degrees of freedom. For
this, modal integral operators are upsampled in a somewhat crude and
costly way. Better adaptivity in this process and in the FFT
operations will lead to increased execution speed.

Nano-optics is another application that involves high-frequency
electromagnetic eigenfields. Here structures that support whispering
gallery modes (WGMs) are of great interest~\cite{Righ11}. The WGMs
have large eigenwavenumbers and large azimuthal indices and the
numerical examples presented in~\cite{HelsKarl15} indicate that our
solver is ideal for their evaluation. When the WGM structures are
dielectric bodies of revolution, the solver needs to be adapted to a
similar set of integral equations as in~\cite{Buly15}. The WGM
structures used in nano-optics are designed to have very high Quality
factors (Q-factors), and the bandwidth of such structures is very
small since the Q-factor equals the frequency-to-bandwidth ratio.
There are reported Q-factors as high as $10^{11}$ \cite{Righ11} with a
relative bandwidth of $10^{-11}$. In principle, a solver that delivers
eleven digit accuracy is needed to design such a structure if a WGM of
the structure is to be excited by a laser with a given wavelength.

\section*{Acknowledgement}

This work was supported by the Swedish Research Council under contract
621-2014-5159.

\appendix

\section{Explicit expressions for kernels}
\label{sec:explicit}

The various double- and single-layer type operators $K_\alpha$ and
$S_\alpha$ used in this paper are defined by their corresponding
static kernels $D_{\alpha}(\vec r,\vec r')$ and $Z_{\alpha}(\vec
r,\vec r')$ via~(\ref{eq:Kdefa}) with~(\ref{eq:Kdefb}) and
(\ref{eq:Sdefa}) with~(\ref{eq:Sdefb}). This appendix collects
explicit expressions for all static kernels along with analytic
expressions for their Fourier coefficients $D_{\alpha n}(r,r')$ and
$Z_{\alpha n}(r,r')$. The numbering of the operators and kernels is
compatible with the numbering used in~\cite{HelsKarl15}. Note that for
azimuthal index $n=0$ the modal operators $K_{2n}$, $K_{3n}$,
$K_{12n}$, $K_{15n}$, $K_{16n}$, $K_{20n}$, $K_{21n}$, $S_{2n}$,
$S_{3n}$, $S_{6n}$, $S_{8n}$, $S_{9n}$, and $S_{13n}$ are zero.

\subsection{Static kernels}
\label{sec:explicitS}

The static kernels are expressed in terms of the azimuthal angle
$\theta$ and vectors and points in the plane defined by $\theta=0$,
see Section~\ref{sec:basic}. The abbreviations
\begin{gather*}
\vec\nu\cdot(\vec r-\vec r')=
\nu\cdot(r-r')+\nu_{\rho}\rho'(1-\cos(\theta-\theta'))\,,\\
\vec\tau\cdot(\vec r-\vec r')=
\tau\cdot(r-r')+\nu_z\rho'(1-\cos(\theta-\theta'))\,,\\
\vert\vec r-\vec r'\vert=\left(\rho^2+\rho'^2 
 -2\rho\rho'\cos(\theta-\theta')+(z-z')^2\right)^\frac{1}{2}\,,\\
\Phi_0(\vec r,\vec r')=\frac{1}{4\pi\vert\vec r-\vec r'\vert}\,,
\qquad
\tilde\Phi_0^3(\vec r,\vec r')=\frac{1}{4\pi\vert\vec r-\vec r'\vert^3}\,,
\end{gather*}
are used. The static double-layer type kernels are
\begin{align*}
D_{\vec\nu}(\vec r,\vec r')&=-\vec\nu\cdot(\vec r-\vec r')
\tilde\Phi_0^3(\vec r,\vec r')\,,\\
D_1(\vec r,\vec r')&= 2\left(\nu_{\rho}'\rho-
(\nu'\cdot r'-\nu_z'z)\cos(\theta-\theta')\right)
\tilde\Phi_0^3(\vec r,\vec r')\,,\\
D_2(\vec r,\vec r')&=-2{\rm i}(z-z')\sin(\theta-\theta')
\tilde\Phi_0^3(\vec r,\vec r')\,,\\
D_3(\vec r,\vec r')&= 2{\rm i}(\nu_z'\nu\cdot r-
\nu_z\nu'\cdot r')\sin(\theta-\theta')
\tilde\Phi_0^3(\vec r,\vec r')\,,\\
D_4(\vec r,\vec r')&=-2\left(\nu_{\rho}\rho'-(\nu\cdot r- 
\nu_zz')\cos(\theta-\theta')\right)
\tilde\Phi_0^3(\vec r,\vec r')\,,\\
D_{11}(\vec r,\vec r')&=\left(\rho-\rho'\cos(\theta-\theta')\right)
\tilde\Phi_0^3(\vec r,\vec r')\,,\\
D_{12}(\vec r,\vec r')&=-{\rm i}\rho'\sin(\theta-\theta')
\tilde\Phi_0^3(\vec r,\vec r')\,,\\
D_{13}(\vec r,\vec r')&=(z-z')
\tilde\Phi_0^3(\vec r,\vec r')\,,\\
D_{14}(\vec r,\vec r')&=
-\vec \tau\cdot(\vec r-\vec r')\nu'_z\cos(\theta-\theta')
\tilde\Phi_0^3(\vec r,\vec r')\,,\\
D_{15}(\vec r,\vec r')&=
 {\rm i}\vec \tau\cdot(\vec r-\vec r')\sin(\theta-\theta')
\tilde\Phi_0^3(\vec r,\vec r')\,,\\
D_{16}(\vec r,\vec r')&=
-{\rm i}\vec \tau\cdot(\vec r-\vec r')\nu'_z\sin(\theta-\theta')
\tilde\Phi_0^3(\vec r,\vec r')\,,\\
D_{17}(\vec r,\vec r')&=-\vec \tau\cdot(\vec r-\vec r')\cos(\theta-\theta')
\tilde\Phi_0^3(\vec r,\vec r')\,,\\
D_{18}(\vec r,\vec r')&=\vec \tau\cdot(\vec r-\vec r')\nu'_\rho
\tilde\Phi_0^3(\vec r,\vec r')\,,\\
D_{19}(\vec r,\vec r')&=
-\vec \nu\cdot(\vec r-\vec r')\nu'_z\cos(\theta-\theta')
\tilde\Phi_0^3(\vec r,\vec r')\,,\\
D_{20}(\vec r,\vec r')&=
 {\rm i}\vec \nu\cdot(\vec r-\vec r')\sin(\theta-\theta')
\tilde\Phi_0^3(\vec r,\vec r')\,,\\
D_{21}(\vec r,\vec r')&=
-{\rm i}\vec \nu\cdot(\vec r-\vec r')\nu'_z\sin(\theta-\theta')
\tilde\Phi_0^3(\vec r,\vec r')\,,\\
D_{22}(\vec r,\vec r')&=-\vec\nu\cdot(\vec r-\vec r')\cos(\theta-\theta')
\tilde\Phi_0^3(\vec r,\vec r')\,,\\
D_{23}(\vec r,\vec r')&=\vec\nu\cdot(\vec r-\vec r')\nu'_\rho
\tilde\Phi_0^3(\vec r,\vec r')\,.
\end{align*}
The static single-layer type kernels are
\begin{align*}
Z_\varsigma(\vec r,\vec r')&=\Phi_0(\vec r,\vec r')\\
Z_1(\vec r,\vec r')&=
 \left(\nu_z\nu_{z}'\cos(\theta-\theta')+\nu_{\rho}\nu_{\rho}'\right)
 \Phi_0(\vec r,\vec r')\,,\\
Z_2(\vec r,\vec r')&=
-{\rm i}\nu_z\sin(\theta-\theta')\Phi_0(\vec r,\vec r')\,,\\
Z_3(\vec r,\vec r')&=
 {\rm i}\nu_z'\sin(\theta-\theta')\Phi_0(\vec r,\vec r')\,,\\
Z_4(\vec r,\vec r')&=
 \cos(\theta-\theta')\Phi_0(\vec r,\vec r')\,,\\
Z_5(\vec r,\vec r')&=
 \left(\nu_\rho\nu_z'\cos(\theta-\theta')-\nu_z\nu_\rho'\right)
 \Phi_0(\vec r,\vec r')\,,\\
Z_6(\vec r,\vec r')&=
-{\rm i}\nu_\rho\sin(\theta-\theta')\Phi_0(\vec r,\vec r')\,,\\
Z_7(\vec r,\vec r')&=\nu'_z\cos(\theta-\theta')\Phi_0(\vec r,\vec r')\,,\\
Z_8(\vec r,\vec r')&={\rm i}\sin(\theta-\theta')\Phi_0(\vec r,\vec r')\,,\\
Z_9(\vec r,\vec r')&=-Z_3(\vec r,\vec r')\,,\\
Z_{10}(\vec r,\vec r')&=Z_4(\vec r,\vec r')\,,\\
Z_{11}(\vec r,\vec r')&=-\nu'_\rho\Phi_0(\vec r,\vec r')\,,\\
Z_{12}(\vec r,\vec r')&=Z_7(\vec r,\vec r')\,,\\
Z_{13}(\vec r,\vec r')&=-Z_8(\vec r,\vec r')\,,\\
Z_{14}(\vec r,\vec r')&=Z_{11}(\vec r,\vec r')\,.
\end{align*}

\subsection{Fourier coefficients}
\label{sec:explicitF}

Our derivation of the Fourier coefficients of the static kernels is
made in a similar manner as in Young, Hao, and
Martinsson~\cite[Section 5.3]{Youn12}. The idea to expand the Green's
function for the Laplacian in the functions
$\mathfrak{Q}_{n-\frac{1}{2}}(\chi)$ comes from Cohl and
Tohline~\cite{Cohl99}. We use the notation of Sections~\ref{sec:basic}
and~\ref{sec:azimuthn} with $\chi$ as in~(\ref{eq:chidef}) and
\begin{align*}
\eta&=\left(8\pi^3\rho\rho'\right)^{-\frac{1}{2}}\,,\\
d(v)&=\frac{v\cdot(r-r')}{|r-r'|^2}\,,\\
\mathfrak{P}_n(\chi)&=\frac{1}{2}\left(\mathfrak{R}_n(\chi)
+\mathfrak{Q}_{n-\frac{1}{2}}(\chi)\right)\,,\\
\mathfrak{M}_n(\chi)&=\chi\mathfrak{Q}_{n-\frac{1}{2}}(\chi)
                      -\frac{\chi+1}{4n^2-1}\mathfrak{R}_n(\chi)\,,\\
\mathfrak{D}_n(\chi)&=\frac{2n(\chi+1)}{4n^2-1}\mathfrak{R}_n(\chi)\,,\\
\mathfrak{L}_n(\chi)&=\mathfrak{R}_n(\chi)+2(\chi-1)\mathfrak{P}_n(\chi)\,,\\
\mathfrak{C}_n(\chi)&=2n(\chi-1)\mathfrak{Q}_{n-\frac{1}{2}}(\chi)\,.
\end{align*}
The Fourier coefficients of the static double-layer type kernels are
\begin{align*}
D_{\vec\nu n}(r,r')&=\eta\left[d(\nu)\mathfrak{R}_n(\chi)
-\frac{\nu_\rho}{\rho}\mathfrak{P}_n(\chi)\right]\,,\\
D_{1n}(r,r')&=-2\eta\left[
d(\nu')\mathfrak{R}_n(\chi)-\frac{\left(\nu'\cdot r'-\nu'_zz\right)}
{\rho\rho'}\mathfrak{P}_n(\chi)\right]\,,\\
D_{2n}(r,r')&=-2\eta
\frac{\left(z-z'\right)}{\rho\rho'}n\mathfrak{Q}_{n-\frac{1}{2}}(\chi)\,,\\
D_{3n}(r,r')&=2\eta
\frac{\left(\nu_z'\nu\cdot r-\nu_z\nu'\cdot r'\right)}{\rho\rho'}
n\mathfrak{Q}_{n-\frac{1}{2}}(\chi)\,,\\
D_{4n}(r,r')&=-2\eta\left[
d(\nu)\mathfrak{R}_n(\chi)+\frac{\left(\nu\cdot r-\nu_zz'\right)}{\rho\rho'}
\mathfrak{P}_n(\chi)\right]\,,\\
D_{11n}(r,r')&=-\eta\left[d(\hat{\rho})\mathfrak{R}_n(\chi)-\frac{1}{\rho}
\mathfrak{P}_n(\chi)\right]\,,\\
D_{12n}(r,r')&=-\frac{\eta}{\rho}n\mathfrak{Q}_{n-\frac{1}{2}}(\chi)\,,\\
D_{13n}(r,r')&=-\eta\,d(\hat{z})\mathfrak{R}_n(\chi)\,,\\
D_{14n}(r,r')&=\eta\,\nu'_z\left[d(\tau)\mathfrak{L}_n(\chi)
-\frac{\nu_z}{2\rho}\left(\mathfrak{L}_n(\chi)+\mathfrak{M}_n(\chi)
\right)\right]\,,\\
D_{15n}(r,r')&=\eta\left[d(\tau)\mathfrak{C}_n(\chi)
-\frac{\nu_z}{2\rho}\left(\mathfrak{C}_n(\chi)+\mathfrak{D}_n(\chi)
\right)\right]\,,\\
D_{16n}(r,r')&=-\eta\,\nu'_z\left[d(\tau)\mathfrak{C}_n(\chi)
-\frac{\nu_z}{2\rho}\left(\mathfrak{C}_n(\chi)+\mathfrak{D}_n(\chi)
\right)\right]\,,\\
D_{17n}(r,r')&=\eta\left[d(\tau)\mathfrak{L}_n(\chi)
-\frac{\nu_z}{2\rho}\left(\mathfrak{L}_n(\chi)+\mathfrak{M}_n(\chi)
\right)\right]\,,\\
D_{18n}(r,r')&=-\eta\,\nu'_\rho\left[d(\tau)\mathfrak{R}_n(\chi)
-\frac{\nu_z}{\rho}\mathfrak{P}_n(\chi)\right]\,,\\
D_{19n}(r,r')&=\eta\,\nu'_z\left[d(\nu)\mathfrak{L}_n(\chi)
-\frac{\nu_\rho}{2\rho}\left(\mathfrak{L}_n(\chi)+\mathfrak{M}_n(\chi)
\right)\right]\,,\\
D_{20n}(r,r')&=\eta\left[d(\nu)\mathfrak{C}_n(\chi)
-\frac{\nu_\rho}{2\rho}\left(\mathfrak{C}_n(\chi)+\mathfrak{D}_n(\chi)
\right)\right]\,,\\
D_{21n}(r,r')&=-\eta\,\nu'_z\left[d(\nu)\mathfrak{C}_n(\chi)
-\frac{\nu_\rho}{2\rho}\left(\mathfrak{C}_n(\chi)+\mathfrak{D}_n(\chi)
\right)\right]\,,\\
D_{22n}(r,r')&=\eta\left[d(\nu)\mathfrak{L}_n(\chi)
-\frac{\nu_\rho}{2\rho}\left(\mathfrak{L}_n(\chi)+\mathfrak{M}_n(\chi)
\right)\right]\,,\\
D_{23n}(r,r')&=-\eta\,\nu'_\rho\left[d(\nu)\mathfrak{R}_n(\chi)
-\frac{\nu_\rho}{\rho}\mathfrak{P}_n(\chi)\right]\,.
\end{align*}
The Fourier coefficients of the static single-layer type kernels are
\begin{align*}
Z_{\varsigma n}(r,r')&=\eta\,\mathfrak{Q}_{n-\frac{1}{2}}(\chi)\,,\\
Z_{1n}(r,r')&=\eta\left[\nu_z\nu'_z\mathfrak{M}_n(\chi)
+\nu_\rho\nu'_\rho\mathfrak{Q}_{n-\frac{1}{2}}(\chi)\right]\,,\\
Z_{2n}(r,r')&=\eta\,\nu_z\mathfrak{D}_n(\chi)\,,\\
Z_{3n}(r,r')&=-\eta\,\nu'_z\mathfrak{D}_n(\chi)\,,\\
Z_{4n}(r,r')&=\eta\,\mathfrak{M}_n(\chi)\,,\\
Z_{5n}(r,r')&=\eta\left[\nu_\rho\nu'_z\mathfrak{M}_n(\chi)
-\nu_z\nu'_\rho\mathfrak{Q}_{n-\frac{1}{2}}(\chi)\right]\,,\\
Z_{6n}(r,r')&=\eta\,\nu_\rho\mathfrak{D}_n(\chi)\,,\\
Z_{7n}(r,r')&=\eta\,\nu'_z\mathfrak{M}_n(\chi)\,,\\
Z_{8n}(r,r')&=-\eta\,\mathfrak{D}_n(\chi)\,,\\
Z_{9n}(r,r')&=-Z_{3n}(r,r')\,,\\
Z_{10n}(r,r')&=Z_{4n}(r,r')\,,\\
Z_{11n}(r,r')&=-\eta\,\nu'_\rho\mathfrak{Q}_{n-\frac{1}{2}}(\chi)\,,\\
Z_{12n}(r,r')&=Z_{7n}(r,r')\,,\\
Z_{13n}(r,r')&=-Z_{8n}(r,r')\,,\\
Z_{14n}(r,r')&=Z_{11n}(r,r')\,.
\end{align*}

\begin{small}

\end{small}

\end{document}